\documentclass{ieeeaccess}
\usepackage{amsmath,amssymb,amsfonts}
\usepackage{graphicx}
\usepackage{textcomp}

\usepackage{enumitem}
\usepackage{subfig}
\usepackage{amsthm}
\usepackage{varioref}

\usepackage[english]{babel}

\usepackage{grffile}
\usepackage{xcolor}
\usepackage{epstopdf}
\usepackage{here}

\usepackage{algorithm}
\makeatletter
\renewcommand{\ALG@name}{Linear program}
\makeatother
\usepackage{algpseudocode}
\usepackage{amsfonts}
\usepackage{siunitx}
\usepackage{xfrac}
\usepackage{mathtools}
\usepackage{units}
\usepackage{eucal}
\usepackage{cancel}

\usepackage{eurosym}
\usepackage{blindtext}
\usepackage{hyperref}

\usepackage{booktabs}
\usepackage{threeparttable}
\usepackage{tablefootnote}
\usepackage{graphicx}

\usepackage[style=ieee,backend=bibtex,bibencoding=ascii,sorting=none,maxbibnames=99]{biblatex}
\usepackage{hyperref,xcolor}
\hypersetup{pdfborder = {0 0 0}}
\bibliography{references}
\AtBeginBibliography{\footnotesize}

\definecolor{mygreen}{rgb}{0.0, 0.5, 0.0}

\def\BibTeX{{\rm B\kern-.05em{\sc i\kern-.025em b}\kern-.08em
    T\kern-.1667em\lower.7ex\hbox{E}\kern-.125emX}}
\begin{document}
\history{Date of submission: Jan 19, 2022 \\ This work has been submitted to the IEEE for possible publication. Copyright may be transferred without notice, after which this version may no longer be accessible.}
\doi{}

\title{Allocation of locally generated electricity in renewable energy communities}
\author{\uppercase{Miguel Manuel de Villena}\authorrefmark{1},
\uppercase{Samy Aittahar}\authorrefmark{1},
\uppercase{Sebastien Mathieu}\authorrefmark{2},
\uppercase{Ioannis Boukas}\authorrefmark{1},
\uppercase{Eric Vermeulen}\authorrefmark{3},
\uppercase{and Damien Ernst}\authorrefmark{1}\authorrefmark{4}
}
\address[1]{Department of Electrical Engineering and Computer Science, University of Liege, Belgium (e-mail: \{mvillena, s.aittahar, ioannis.boukas, dernst\}@uliege.be)}
\address[2]{delpower, Belgium (e-mail: sebastien.mathieu@delpower.be)}
\address[3]{haulogy, Belgium (e-mail: eric.vermeulen@haulogy.net)}
\address[4]{LTCI, Telecom Paris, Institut Polytechnique de Paris}
\tfootnote{This research has been supported by haulogy and the Public Service of Wallonia in the framework of the haulogy 2021 R\&D programme launched by haulogy which publishes software for energy communities.}

\markboth
{M. Manuel de Villena \headeretal: Allocation of locally generated electricity in renewable energy communities}
{M. Manuel de Villena \headeretal: Allocation of locally generated electricity in renewable energy communities}

\corresp{Corresponding author: Miguel Manuel de Villena (e-mail: mvillena@uliege.be).}

\begin{abstract}
	Local electricity markets represent a way of supplementing traditional retailing contracts for end consumers---among these markets, the renewable energy community has gained momentum over the last few years.
	This paper proposes a practical and readily to be adopted modelling solution for these communities, one that allows their members to share the economic benefits derived from them.
	The proposed solution relies on an \emph{ex-post} allocation of the electricity that is generated within energy communities (i.e., local electricity) based on the optimisation of \emph{repartition keys}.
	Repartition keys are therefore optimally computed to represent the proportion of total local electricity to be allocated to each community member, and aim to minimise the sum of electricity bills of all community members.
	Since the optimisation takes place \emph{ex-post} the repartition keys do not modify the actual electricity flows, but rather the financial flows of the community members.
	Then, the billing process of the community will take these keys into account to correctly send the electricity bills to each member.
	Building on this concept, we also introduce two additions to the basic algorithm to enhance the stability of the community, which a global bill minimisation may fail to ensure (e.g., very asymmetrical solutions between members may lead to some of them opting out).
	The first addition is the computation of the self-sufficiency rates of the community members, defined as the proportion of the electricity demand covered by local electricity---this can be used to ensure a more even allocation of the local electricity among the community members, effectively acting as a revenue sharing system.
	The second addition is the use of initial repartition keys based on which the optimised ones are computed, and a tolerance on the maximum deviations between both sets of keys---this can help decrease the uncertainty of potential community members prior to their participation, as it ensures a minimum---contractual---level of revenue for each of them.
	We have tested our initial methodology with a broad range of scenarios illustrating its ability reduce the electricity bills of the community members.
	Likewise, the two additions are tested showcasing the impact of revenue sharing by means of self-sufficiency rates and initial keys.
	Our results show that creating a community using this methodology can potentially reduce the electricity costs for all community members, and that self-sufficiency rates and initial keys can be used to stabilise the community by performing revenue sharing among them.
\end{abstract}

\begin{keywords}
    Distributed generation, energy communities, local electricity market, repartition keys, revenue sharing.
\end{keywords}

\titlepgskip=-15pt

\maketitle

\section*{Notation}

\begin{tabbing}
xxxxxxxxx\= xxxxxxxxxxxxxxxxxxxxxxxxxx \kill
\textit{Sets} \\
$\mathcal{T}$\> Set of market periods $\left\{ 1, \ldots, T \right\}$ \\
$\mathcal{I}$\> Set of REC members $\left\{ 1, \ldots, I \right\}$ \\
\\
\textit{Parameters} \\
$A_{t,i}$\> Initial allocation of electricity generation \\
$C_{t,i}$\> Consumption \\
$C^n_{t,i}$\> Netted (after-the-meter) consumption \\
$K_{t,i}$\> Initial repartition keys \\
$P_{t,i}$\> Electricity generation \\
$P^n_{t,i}$\> Netted (after-the-meter) electricity generation \\
$SSR_{i}^{min}$\> Minimum self-sufficiency rate \\
$X_{t,i}$\> Maximum allowed key deviation \\
$\xi_{i}^{b}$\> Retail price of electricity \\
$\xi_{i}^{s}$\> Selling price of electricity to the grid \\
$\xi_{i}^{l-}$\> Purchasing price of electricity from the REC \\
$\xi_{i}^{l+}$\> Selling price of electricity to the REC \\
$\xi_{i}^{d}$\> Price of deviations from $A_{t,i}$ \\
\\
\textit{Decision variables} \\
$a_{t,i}$\> Optimised allocated production \\
$\overline{a}_{t}^{+}$\> Positive deviation of allocated generation from $A_{t,i}$ \\
$\overline{a}_{t}^{-}$\> Negative deviation of allocated generation from $A_{t,i}$ \\
$k_{t,i}$\> Optimised repartition keys \\
$ssr_{i}$\> Self-sufficiency rate \\
$v_{t,i}$\> Verified allocated electricity generation \\
$y_{t,i}$\> Locally sold electricity generation
\end{tabbing}

\section{Introduction}
\label{sec:introduction}
\PARstart{U}{nder} the pressure imposed by climate change and global warming, decarbonising the electricity sector has become one of the key targets of policies and regulations worldwide---in Europe, the Clean Energy Package for all Europeans \cite{EUcleanpackage} and the UK Climate Act \cite{UKclimateact}, demonstrate the commitment of the region toward this target. 
Decarbonising the electricity sector requires the widespread investment in renewable electricity generation resources, and traditionally can be achieved through a centralised or a decentralised investment endeavour \cite{bouffard2008centralised,manueldevillena2021smart}.
According to the latter, multiple small electricity generation units, usually owned by final consumers, are deployed near or directly at the consumption centres (e.g., rooftop solar panels).
In recent years, this decentralised approach has gained momentum, owing to various technological advances in generation and storage technologies, automated energy management systems, or trading bots used for the financial optimisation of these generation units \cite{parag2016electricity,shinde2021analyzing}.
This type of approach, however, demands for regulatory frameworks that effectively address the new needs of final consumers, who can now be electricity generation owners (i.e., prosumers).
Indeed, the lack of adequate frameworks for consumers and prosumers challenges the power system operation, as shown in \cite{schittekatte2020least,abada2020viability,manuelde2021network,manuelde2021modelling}, and may hinder the integration of decentralised generation \cite{heaslip2016assessing}.
A potential solution to this problem is the creation of local consumer-centric electricity markets, where consumers and prosumers can trade electricity locally, replacing partial or totally their traditional retailer contracts.
In this regard, one type of local market that has received considerable attention over the last few years is the renewable energy community (REC).
RECs represent, according to \cite{sokolowski2018european}, a cost-efficient way to meet the energy needs of citizens---they allow their participants (REC members) to exchange among them the electricity generated by their own generation assets.
Despite their recent popularity, the rules controlling the internal electricity exchanges within RECs are not yet clearly defined, which may prevent their rollout \cite{reijnders2020energy}.
This paper aims to fill this gap, proposing a methodology which is compliant with current European regulations and can be readily implemented in practice.

According to European regulations, RECs are composed of final customers (consumers and prosumers) which can exchange electricity among them through a local market \cite{union2018directive}.
As per this directive, consumers are entitled to participate in an REC without losing their rights as final customers (e.g., they may freely choose their electricity supplier), and are centrally managed by an energy community manager (ECM).
One of the key roles of the ECM is to facilitate the emergence and stability of the internal market.
To that end, the ECM must ensure that the local electricity generation is optimally allocated among the REC members so as to meet some objective, such as the minimisation of the sum of their electricity bills.
Various methodologies may be applied in this regard, and the existing literature on optimisation techniques can provide solutions to this class of problems.

\subsection{Literature review}

The internal electricity exchanges between the members of a consumer-centric market (e.g., microgrids, smart grids, virtual power plants, or RECs) may be modelled utilising various market mechanisms.
Parag and Sovacool \cite{parag2016electricity} propose three types of such mechanisms: peer-to-peer (P2P), prosumer-to-grid, and organised groups of prosumers.
In the context of RECs, both P2P and prosumer groups are suitable approaches, and thereby models considering these two options can offer useful insights into the establishing of these communities.

One of the first papers introducing the idea of P2P decentralised trading dates back to more than 20 years \cite{wu1999coordinated}, and models the negotiations between agents depending on their cost functions.
Since then, several studies have focused on algorithms for P2P exchanges in a decentralised fashion using different techniques based either on bilateral trading where pairs of market participants negotiate the electricity transactions independently, or on the idea of cooperative or non-cooperative games between all the players \cite{tushar2018transforming,mengelkamp2018designing,ceer2019,caramizaru2020energy,hahnel2020becoming}.
A multi-bilateral economic dispatch method where peers are free to negotiate with any other peer in their network (i.e., disregarding any potential network constraint) is introduced by Sorin et al. \cite{sorin2018consensus}, and later compared with other approaches by Moret et al. \cite{moret2018negotiation}.
Another market design based on bilateral contract networks for P2P trading is presented by Morstyn et al. \cite{morstyn2018bilateral}---this new design stands out by introducing real-time and forward (e.g., day-ahead) markets.
Some studies have provided solutions based on game theoretic methodologies. 
For example, using a Nash equilibrium non-cooperative game, Zhang et al. \cite{zhang2018peer} show that P2P trading within a microgrid leads to increased self-sufficiency rates and reduced peak loads in the overall microgrid, thereby fostering the integration of decentralised generation technologies.
Paudel et al. \cite{paudel2018peer} apply a multi-game methodology in real-time to model the exchanges between prosumers in a microgrid---two consecutive games are employed to determine the prices offered by sellers, and the matching between buyers and sellers.
Wei et al. \cite{wei2021optimal} make use of robust optimisation to account for wind power uncertainty and a Nash bargaining problem to promote P2P trading in a microgrid context---the problem is then solved with an augmented Lagrangian multipliers method.
Different papers deal with coalitional game in the context of P2P exchanges, whereby groups of consumers (coalitions) may form, improving their overall cost function, and then share the payoff among them.
In \cite{tushar2019motivational}, Tushar et al. employ motivational psychology to demonstrate the suitability of coalitional game theory to model P2P trading in the context of smart grids.
This coalitional framework is further explored and developed in \cite{tushar2020coalition} by the same authors.
Also employing a coalitional game, Safdarian et al. \cite{safdarian2021coalitional} simulate an REC in an apartment building where a clustering technique groups the consumers in different clusters to attain savings in their electricity costs.
Tushar et al. \cite{tushar2018transforming} extensively address the topic of P2P energy management using game theory, comparing several models for managing P2P networks and highlighting the main criteria to be considered when designing such problems.
Finally, very few studies have taken into account the role of the distribution system operator (DSO) when determining the exchanges, which may lead to suboptimal solutions.
In this regard, Chen et al. \cite{chen2021peer} propose a P2P approach applied to local groups of prosumers with a dynamic network structure controlled by the DSO, leading to significant network losses reductions.

Although all these approaches provide useful methods to sharing electricity among consumers and prosumers, they fail to account for their potential coordination through, for instance, a central planner.
This lack of central coordination may lead to welfare losses due to (i) hidden information and (ii) possible conflicting objectives among the different actors.
Hidden information may induce overall welfare losses by restricting the feasible space of solutions to the information that is shared.
Moreover, conflicting objectives may result in suboptimal solutions for all prosumers in the case of a decentralised approach due to differing goals among the market participants.
Whilst differing objectives may also have an impact using a centralised paradigm, a central planner will yield a pareto-optimal front of optimal solutions, leading to trade-offs among participants, which is not the case for a fully decentralised optimisation problem.

On the other hand, adopting a centralised planning approach to handle the operation of microgrids or RECs may solve some of these issues.
Several authors have studied the problem of decentralised trading in the context of these types of local market structures.
Long et al. \cite{long2017peer}, Liu et al. \cite{liu2017energy}, and Grzani{\'c} et al. \cite{grzanic2021electricity} focus on the internal electricity market and how to compute the internal prices (i.e., prices associated to the internal exchanges), comparing different price mechanisms \cite{long2017peer,grzanic2021electricity} and introducing demand response \cite{liu2017energy,grzanic2021electricity}.
From a different standpoint, Hu et al. \cite{hu2021multi} analyse the possibility of microgrid market participation, including balancing as well as ancillary markets.
Soriano et al. \cite{soriano2021peer} and Mashlakov et al. \cite{mashlakov2021decentralized} make use of multiobjective optimisation techniques to model the electricity exchanges between prosumers.
Faqiry and Das \cite{faqiry2016double}, propose a double-sided auction mechanism in the context of islanded microgrids where buyers post bids to buy electricity and sellers offer their availabilities.
The market is cleared by an aggregator that, iteratively, converges to the market clearing price.
In \cite{moret2018energy}, Moret et al. present an REC where the ECM acts as the interface between the REC members and the market---members do not interact with their retailers but rather with the ECM, who supplies their electricity, computes the electricity prices and the internal exchanges, and charges them.
Morstyn et al. \cite{morstyn2018using} and Sousa et al. \cite{sousa2019peer} discuss different consumer-centric market mechanisms: P2P trading, community trading (termed virtual power plant trading in \cite{morstyn2018using}) and a hybrid approach mixing both mechanisms.
Both studies conclude that the hybrid approach may create an overall more valuable and scalable solution.
Another approach based on central planning is presented by Corn{\'e}lusse et al. \cite{cornelusse2019community}, where a benevolent central planner maximises the welfare of the community and redistributes revenues and costs among the REC members so that none of them is penalised as a result of being in a community.
This problem is cast as a bi-level optimisation where the lower level solves the clearing problem of the community and the upper level shares the profits among the entities.
In \cite{manueldevillena2020}, Manuel de Villena et al. consider a problem where flexibility from prosumers can be activated by the central planner to minimise electricity costs.
Two approaches based on cooperative game theory \cite{abada2020viability,mitridati2021design} analyse the value and viability of RECs from a centralised standpoint.
Abada et al. \cite{abada2020viability} propose the use of a sharing rule for REC members based on the gains stemming from local production and consumption, as opposed to only production as it is usually done.
Finally, Mirtriadi et al. \cite{mitridati2021design} create a cooperative game where the values (welfare of REC members) obtained through different allocation mechanisms are compared.
The presented studies consider some central agent to act as the interface between electricity markets and end consumers and to coordinate the P2P exchanges---all of them concur that P2P trading through a central agent may unlock significant savings in energy costs for the participants of the microgrid or energy community.

\subsection{Contributions of this work}

Although all the analysed studies provide sound frameworks to assess the economic benefits of internal trading in the context of microgrids or energy communities, their potential implementation in real-life applications is limited.
There are two main challenges hindering their implementation: (i) complexity, as real energy communities may comprise many different agents which may not always be adequately clustered, scalability problems quickly emerge where the problem becomes NP-hard and thereby a trade off between number of participants and reasonable solving time is needed; and (ii) lack of adherence to the regulatory frameworks in place, which may limit, for instance, the role of the ECM.
A possible option to overcome the first challenge is to perform an \emph{ex-post} computation of electricity prices and exchanges.
This would imply that, after the physical delivery of electricity, there is a phase where the financial exchanges according to which the REC members are billed can be determined.
To apply this approach, a sharing rule among REC members can be defined, but the actual exchanges between them do not need to be negotiated, guaranteeing the fast convergence of the method and making the problem relatively simple to solve.
To the best of our knowledge, in the current literature only Grzani{\'c} et al. \cite{grzanic2021electricity} makes use of such an approach.
In their work, the authors model a cost-sharing mechanism among the prosumers of an REC relying on the assumption that the REC is managed by an ECM, which is also the electricity supplier and organiser of the internal market.
Such an assumption, though, may conflict with the second challenge (adherence to regulatory frameworks).
Indeed, according to the definition of RECs by the European Commission, this methodology would violate the right of REC members to freely choose their electricity supplier.

Against this backdrop, our work proposes an approach which is practical, scalable, relatively simple to deploy in real RECs, and compliant with current regulations\footnote{In this paper we focus on the current legislative framework laid out by the European Commission \cite{union2018directive}, with regards to the role of the ECM. However, the problem is designed so that it can readily be adapted to other roles, such as the aggregation or supply of the REC members.}.
To that end, we have designed a methodology, based on the concept of \emph{repartition keys}, to determine the electricity exchanges between the REC members of a centrally managed REC.
The use of repartition keys, although not conventional, is not new, and it is, in fact, foreseen by the French \cite{code2017d3156} as well as the Walloon (Belgium) regulations \cite{spw2019communautes}.
Moreover, other European countries are adopting similar legislative decisions \cite{torabi2020mainstreaming}.
In our particular case, repartition keys represent the proportion of locally generated electricity, measured after-the-meter (i.e., after meeting the electricity needs of the prosumer who generates it), which is allocated to each REC member---there is one repartition key per member.
These keys are optimally computed by the ECM through an optimisation framework aiming to minimise the sum of electricity bills of the REC members.
To guarantee the scalability and convergence of this approach, this optimisation problem is solved \emph{ex-post} utilising a centralised approach.
This way of distributing the available local electricity has the advantage of being a single centralised computation, which decreases the complexity of the problem as REC members do not need to negotiate the electricity trades among them.
Moreover, this approach may lead to stability issues whereby REC members receive asymmetrical repartition keys (i.e., asymmetrical fractions of the total locally generated electricity), which may eventually lead to the departure of some members from the REC as they do not receive enough benefits from their participation.
To tackle this pitfall, the problem of stability is analysed in depth in our work, making sure that trade-offs can be found through our optimisation problem.
According to this optimisation problem, REC members' electricity demands are partial or totally met by the surplus of other REC members, which is allocated to them through the repartition keys.
If the electricity provided locally is insufficient to fully cover their needs, REC members can resort to traditional retailer contracts and, to comply with current regulation, they are free to choose their electricity supplier.
Consequently, the electricity demand of each REC member is split into two flows: (i) the proportion of the overall local production allocated to them through repartition keys; and (ii) the electricity supplied by each member's electricity supplier.
The sum of those flows must amount to the total electricity consumption of each REC member.

Since the computation of the repartition keys takes place \emph{ex-post}, the actual power flows are not affected by it, similar to \cite{grzanic2021electricity}.
However, whilst in their work the authors propose a market clearing mechanism organised by the ECM (which is also the only supplier of the REC) that accounts for different cost-sharing rules, in our work this is not possible insofar as the REC members must have their own suppliers.
Indeed, in our approach, the repartition keys are optimised by a central entity (ECM) and communicated to the DSO so that the meter readings of the REC members can be modified accordingly (smart meters are therefore required).

Following this introduction, the remainder of the paper is structured as follows: Section~\ref{sec:rec_and_repartition_keys} formalises the allocation problem faced by the ECM to compute repartition keys that distribute the electricity locally generated within an REC. Section~\ref{sec:rec_sustainability} discusses the stability and sustainability of the REC and proposes two alternative formulations that have the potential to make the REC more stable. Section~\ref{sec:case_study} introduces four examples that illustrate the proposed algorithms. Finally, Section~\ref{sec:conclusion} concludes the paper.

\section{Energy community with repartition keys}
\label{sec:rec_and_repartition_keys}

This section presents the problem of computing the electricity exchanges between the members of an REC.
In particular, we describe this as an allocation problem whereby the electricity produced by the REC members which cannot be absorbed by them, is made available for the rest of the REC members.
The ECM must, therefore, optimally distribute this available electricity among the REC members.
To carry out this optimisation, in this paper we put forward an aggregated cost minimisation of the sum of electricity bills of all REC members.
These bills, charged to them by their electricity suppliers, consist of four elements: (i) imports from the supplier, (ii) exports to the supplier; (iii) imports from the REC, and (iv) exports to the REC.

As previously explained, these four flows can be controlled through repartition keys representing the percentage of total available local electricity assigned to each REC member.
The repartition keys are optimally computed \emph{ex-post} (i.e., after the physical delivery of electricity) by the ECM, whose access to the time-series of demand and generation must be granted---one repartition key per REC member and metering period (defined as the meter's resolution e.g., 15 minutes) is computed.
Performing this computation \emph{ex-post} guarantees the scalability and convergence of our approach, and facilitates the flow of information since the demand and generation are already known (i.e., no need to rely on forecasts).
The new flows obtained after the computation of the repartition keys (imports and exports from/to the supplier and the REC) are then relayed to the DSO managing the meters so that the meter readings of the REC members can be modified accordingly.
Each REC member's electricity supplier can then compute the electricity bill accounting for the optimisation of the keys.

\subsection{Mathematical definition of the REC with repartition keys}
\label{subsec:problem_statement}

Let $\mathcal{T} = \left\{ 1, \ldots, T \right\}$ denote the set of all metering periods in the optimisation horizon where $T$ is the last time step of the simulation's horizon.
The metering period is defined by the intervals $\left( t, t+1 \right]$ contained in $\mathcal{T}$.
In addition, a set of REC members is defined as $\mathcal{I} = \left\{ 1, \ldots, I \right\}$.
These members are characterised by their total production --if any-- and consumption profiles, given as time series whose resolution is equal to the metering period, and span the entire optimisation horizon.
Since REC members may be prosumers, their consumption per metering period must be netted so that their electricity generation is first used to cover their own electricity demand.
The consumption and net consumption are denoted by $C_{t,i}$ and $C_{t,i}^{n}$, respectively.
Similarly, the production must be netted to account for any behind-the-meter consumption.
The production and net production are denoted by $P_{t,i}$ and $P_{t,i}^{n}$, respectively.
\begin{gather}
	\label{eqn:net_consumption}
	C_{t,i}^{n} =  \max \left\{ 0, C_{t,i} - P_{t,i} \right\} \quad \forall (t,i) \in \mathcal{T} \times \mathcal{I}, \\
	\label{eqn:net_production}
	P_{t,i}^{n} = \max \left\{ 0, P_{t,i} - C_{t,i} \right\} \quad \forall (t,i) \in \mathcal{T} \times \mathcal{I}.
\end{gather}

Commonly, producers sell some of their net production to the community, which may not be able to consume it all.
The local production sold by REC member $i$ at metering period $\left( t, t+1 \right]$ is denoted as $y_{t,i}$ and bounded by $P_{t,i}^{n}$, as:
\begin{gather}
	\label{eqn:bound_sold_production}
	y_{t,i} \leq P_{t,i}^{n} \quad \forall (t,i) \in \mathcal{T} \times \mathcal{I}.
\end{gather}

The set of repartition keys are defined per REC member $i$ and metering period $t$.
They are computed through the decision variables $k_{i,t}$ and can be used to determine the allocated production to each REC member at each metering period, such that:
\begin{gather}
	\label{eqn:optimal_allocated_production}
	a_{t,i} = k_{t,i} \cdot \sum_{i \in \mathcal{I}} P^n_{t,i} \quad \forall (t,i) \in \mathcal{T} \times \mathcal{I},
\end{gather}
where $a_{t,i}$ represents the allocated production.
This allocated production, nevertheless, is not necessarily the one accepted by the DSO to correct the meter readings.
For instance, if the total net production ($\sum_{i \in \mathcal{I}} P_{t,i}^{n}$) is greater than the total net consumption ($\sum_{i \in \mathcal{I}} C_{t,i}^{n}$), Equation~\eqref{eqn:optimal_allocated_production} may lead to allocations that are, in fact, larger than the total net consumption.
To avoid such situations, a second set of constraints computes the verified allocated production $v_{t,i}$, which takes the smaller value between the allocated production and the net consumption:
\begin{gather}
	\label{eqn:validated_allocated_production}
	v_{t,i} = \min \left\{ a_{t,i}, C^n_{t,i} \right\} \quad \forall \left( t,i \right) \in \mathcal{T} \times \mathcal{I}.
\end{gather}
In addition, the sum of the verified allocated production must be equal to the sum of local production sold over the set $\mathcal{I}$, for each metering period:
\begin{gather}
	\sum_{i \in \mathcal{I}} v_{t,i} = \sum_{i \in \mathcal{I}} y_{t,i} \quad \forall t \in \mathcal{T}.
\end{gather}

The computation of the repartition keys aims at minimising the sum of individual electricity bills of the REC members, expressed as:
\begin{gather}
	\label{eqn:billing_costs}
	\begin{split}
		B_{t,i} = &\xi_{i}^{b} \cdot \left( C^n_{t,i} - v_{t,i} \right) + \xi_{i}^{l-} \cdot v_{t,i} - \\
		&\xi_{i}^{l+} \cdot y_{t,i} - \xi_{i}^{s} \cdot \left( P^n_{t,i} - y_{t,i} \right) \quad
	\end{split}
	\forall (t,i) \in \mathcal{T} \times \mathcal{I}, 
\end{gather}
where $\xi_{i}^{b}$ is the overall retail electricity price of member $i$ (including transmission, distribution, commodity, and taxes), and $\xi_{i}^{s}$ is the electricity selling price to the supplier of member $i$.
Similarly, $\xi_{i}^{l-}$ is the internal electricity price for imports including taxes, local distribution (which may also include a fee for the transmission system operator), and energy, and $\xi_{i}^{l+}$ is the internal selling price of electricity.

\subsection{Computation of the repartition keys}
\label{subsec:problem_formulation}

The problem of allocating locally generated production by means of repartition keys can be expressed as a linear program.

\begin{algorithm}
\label{lp:lp1}
\caption{Allocation of local electricity through repartition keys}

\begin{equation}
	\label{eqn:objective_function}
	\displaystyle\min_{z \in Z}
	\sum_{i \in \mathcal{I}} \sum_{t \in \mathcal{T}} B_{t, i},
\end{equation}
subject to:
\begin{gather}
	\label{eqn:keys_computation}
	a_{t,i} = k_{t,i} \cdot \sum_{i \in \mathcal{I}} P^n_{t,i} \quad \forall (t,i) \in \mathcal{T} \times \mathcal{I}, \\
	\label{eqn:allocation_limit}
	\sum_{i \in \mathcal{I}} v_{t,i} = \sum_{i \in \mathcal{I}} y_{t,i} \quad \forall t \in \mathcal{T}, \\
	\label{eqn:sold_production_limit}
	y_{t,i} \leq P^n_{t,i} \quad \forall (t,i) \in \mathcal{T} \times \mathcal{I}, \\
	\label{eqn:verified_allocation_limit1}
	v_{t,i} \leq a_{t,i} \quad \forall (t,i) \in \mathcal{T} \times \mathcal{I}, \\
	\label{eqn:verified_allocation_limit2}
	v_{t,i} \leq C^n_{t,i} \quad \forall (t,i) \in \mathcal{T} \times \mathcal{I}, \\
	\label{eqn:share_keys}
	\sum_{i \in \mathcal{I}} k_{t,i} \leq 1 \quad \forall t \in \mathcal{T} \\
    \label{eqn:limit_k}
	k_{t,i} \in \left[ 0, 1 \right], \\
    \label{eqn:limit_vars}
	y_{t,i}, a_{t,i}, v_{t,i} \in \mathbb{R}_{+}.
\end{gather}

\end{algorithm}

In this linear program, the vector of decision variables is $z = \left( k_{t,i}, y_{t,i}, a_{t,i}, v_{t,i} \right) \in Z$.
Note that there is no time-coupling in this linear program and that the objective function is aggregated over $\mathcal{T}$ for simplicity to reflect the real electricity bills---the linear problem could be solved $T$ times and then aggregated to obtain the same results.

The objective function (Equation~\eqref{eqn:objective_function}) aims at minimising the sum of electricity bills of the REC members (see Equation~\eqref{eqn:billing_costs}.

Equation~\eqref{eqn:keys_computation} computes the optimised allocated production.
Equation~\eqref{eqn:allocation_limit} sets the total allocated production equal to the total production sold by the REC members. 
Equation~\eqref{eqn:sold_production_limit} limits the production sold to the total available production. 
Equations~\eqref{eqn:verified_allocation_limit1}~and~\eqref{eqn:verified_allocation_limit2} limit the verified allocated production to the smaller value between allocated production and demand. 
Finally, Equation~\eqref{eqn:share_keys} limits the sum of the repartition keys of the REC members to $1$ (100\%), meaning that the local production shared among the REC members cannot exceed the total electricity produced within the REC.

\section{Sustainability of the community}
\label{sec:rec_sustainability}

The optimisation of repartition keys through a central manager such as the ECM leads to the minimum overall costs for the REC.
Through this optimisation problem, outlined in Equations~\eqref{eqn:objective_function}~--~\eqref{eqn:share_keys}, the ECM distributes the total available local electricity among the REC members depending on the associated price signals for each type of electricity exchange. First, the demand of the REC members is covered with the local generation available, assuming that REC internal exchanges are less expensive than that of the suppliers. Then, if the available local generation is not enough, the REC members purchase the remaining of their electricity needs from their electricity suppliers; if, on the other hand, the available local generation is in excess, the REC members sell the surplus to their electricity suppliers.
As stated in the introduction, this optimisation problem is solved as a single centralised computation, decreasing the complexity of the problem since REC members do not need negotiate the electricity trades---the resulting problem is computationally inexpensive, as demonstrated in Appendix~\ref{app:complexity_analysis}.
Moreover, since the optimisation problem can be cast as a linear program, the minimum costs obtained are globally optimal.

This approach though, is not free of challenges.
Indeed, the global optimisation will find a global optimum (minimum costs) for the whole REC, however, the individual costs of each REC member are not necessarily minimised.
For instance, if two REC members have differing retailer contracts, the centralised algorithm will allocate the local electricity to the member with the pricier contract first, and only then, allocate it to the cheaper one.
In consequence, some REC members may not perceive enough overall benefits from the REC, which has the potential to destabilise the REC as some of the members may leave the energy community\footnote{Note that, although in the context of our work we do not analyse the overhead of becoming part of an REC, it is foreseeable that REC members incur in some extra costs for participating (e.g., ECM fee)}.

This is a known problem, and several authors in the existing literature have described and tackled it by resorting to cost and revenue sharing among the REC members through a market clearing mechanism, ensuring that no REC member is penalised as a result of their participation in an REC \cite{long2017peer,grzanic2021electricity,moret2018energy,cornelusse2019community}.
However, these approaches demand that the ECM be not only the market facilitator but also the electricity supplier of all REC members, which is not foreseen by the current regulation.
To overcome this pitfall, we propose two alternative solutions to the cost and revenue sharing problem based on: (i) the computation of self-sufficiency rates; and (ii) the use of initial repartition keys which are exploited in the optimisation problem to compute the optimal repartition keys.
In the remainder of this section, the two options are discussed.

\subsection{Revenue sharing through self-sufficiency rate}
\label{subsec:ssr}

The first of our proposed solutions to formulate a cost and revenue sharing mechanism where the ECM does not need to act as the electricity supplier of the REC, is based on the computation and optimisation of the self-sufficiency rate (SSR).
The SSR is defined as the fraction of the REC members' electricity consumption which is covered by local electricity generation (i.e., within the REC), and can be computed on an aggregated and on an individual basis (i.e., for the entire REC and for each REC member individually).
In the following, the definition of the SSR for one REC member is presented.

The covered consumption of member $i$ is equal to the sum of the member $i$'s self-consumption and the local electricity generation from other members allocated to member $i$: this is calculated as $P_{t,i} - y_{t,i} + v_{t,i}$.
However, since the allocated production may be greater than the total consumption $C_{t,i}$, the covered consumption must be expressed as $\min\left\{ P_{t,i} - y_{t,i} + v_{t,i}, C_{t,i} \right\}$.
In this last expression, if $y_{t,i}$ is positive, then $P_{t,i} - y_{t,i} + v_{t,i}$ must be greater or equal than $C_{t,i}$ by design of the optimisation problem (see Equations~\eqref{eqn:net_production}~and~\eqref{eqn:bound_sold_production}).
Therefore, the previous expression can be simplified as $\min\left\{ P_{t,i} + v_{t,i}, C_{t,i} \right\}$.
Formally, the SSR over the horizon $T$ of member $i$ is denoted by $ssr_{i}$ and computed as:
\begin{equation}
	\label{eqn:self_sufficiency_rate_def}
	ssr_{i} = \dfrac{\sum_{t \in \mathcal{T}} \min\left\{ P_{t,i} + v_{t,i}, C_{t,i} \right\}}{\sum_{t \in \mathcal{T}} C_{t,i}} \quad \forall i \in \mathcal{I}.
\end{equation}

According to this definition, the SSR depends on the amount of local electricity each REC member receives from the REC and on their own consumption.
Consequently, assuming that the internal exchanges within the REC are less costly than the traditional retailer contract of the REC members, the SSR strongly relates to the potential cost savings of the REC members.
If this assumption holds true, the SSR is a direct measure of the level of benefit each REC member derives from the REC.
In that case, the SSR can be used to ensure that all REC members benefit from their participation in the REC in terms of energy and cost.
To that end, a minimum SSR, denoted by $SSR_{i}^{min}$ may be defined and enforced to each member as:
\begin{gather}
	\label{eqn:min_self_sufficiency_rate_def}
	SSR_{i}^{min} \leq ssr_{i} \quad \forall i \in \mathcal{I},
\end{gather}
therefore enduring a minimum return to each participant joining the energy community.

The concept of minimum SSR can be integrated in the minimisation algorithm presented in the previous section, embedding it in the optimisation problem.
Thus, a second linear program can be written:

\begin{algorithm}
\label{lp:lp2}
\caption{Computation of repartition keys with minimum SSR.}

\begin{equation}
	\label{eqn:objective_function2}
	\displaystyle\min_{z \in Z}
	\sum_{i \in \mathcal{I}} \sum_{t \in \mathcal{T}} B_{t, i},
\end{equation}
subject to:
\begin{gather}
	\nonumber \text{Equations \eqref{eqn:keys_computation} -- \eqref{eqn:limit_vars}}, \\
	\label{eqn:self_sufficiency_rate}
	SSR_{i}^{min} \leq \dfrac{\sum_{t \in \mathcal{T}} \min\left\{ P_{t,i}, C_{t,i} \right\} + v_{t,i} }{\sum_{t \in \mathcal{T}} C_{t,i}} \quad \forall i \in \mathcal{I}.
\end{gather}

\end{algorithm}

In this linear program, the vector of decision variables is $z = \left( k_{t,i}, y_{t,i}, a_{t,i}, v_{t,i} \right) \in Z$.

The new Equation~\eqref{eqn:self_sufficiency_rate} computes the SSR of every member ($ssr_{i}$), enforcing minimum values ($SSR_{i}^{min}$) for all of them.
Note that the enforced values are encoded as a vector of minimum SSR of size $|\mathcal{I}|$ (one per member), therefore each member may have a different enforced value.
This last equation may lead to infeasible solutions by enforcing overly ambitious $SSR_i^{min}$---the SSR depends on the real consumption and generation within the REC and Equation~\eqref{eqn:self_sufficiency_rate} only redistributes it among the REC members.
If the minimum SSR enforced (i.e., $SSR_{i}^{min}$) leads to an unfeasible solution, it needs to be redefined by the ECM.
Observe that the numerator in Equation~\eqref{eqn:self_sufficiency_rate} is a linearisation of the numerator in Equation~\eqref{eqn:self_sufficiency_rate_def}.
The two versions can be shown to be equivalent, and the mathematical proof can be found in Appendix~\ref{app:linearisation_ssr}.

\subsection{Minimum revenue through initial repartition keys}
\label{subsec:initial_keys}
The second solution we propose to the formulation of a cost sharing mechanism that helps ensure the sustainability of the REC, without the need for the ECM to act as the electricity supplier of all REC members, makes use of a set of initial repartition keys in addition to the optimised ones.
With these two sets, the optimal keys would be computed based on the initial ones.
Introducing a set of initial repartition keys can help reduce the uncertainty faced by REC members when joining an REC---as we will see in this section, the initial keys can help ensure a minimum revenue per REC member, increasing the stability of the REC as members are more certain of the outcome of their participation.

This initial set of keys should be contractual (i.e., agreed upon between the ECM and the REC members) and, similarly to the optimised set of keys, represent an agreed --initial-- allocation of the local electricity generation to be provided to each REC member.
The set of initial keys can be pre-computed by the ECM based on historical consumption data of the prospective members, as well as historical data on total available electricity allocation.
Then, based on the initial keys, the ECM can recalculate (optimise) the allocation of local electricity generation, computing the optimal set of repartition keys.
When performing this optimisation, a tolerance can be agreed and set around the initial repartition keys so that the optimised ones do not deviate beyond some hard limit.
As previously explained, the set of initial keys and a the tolerance reduce the uncertainty faced by REC members toward the decision to join the REC, as they have prior knowledge on the minimum revenue the will perceive (i.e., agreed initial keys minus tolerance), thus increasing the probability of a successful and sustainable community.

Let $K_{i}$ denote the set of initial repartition keys, which is only member dependent (i.e., a vector of keys whose size is $|\mathcal{I}|$).
These keys lead to an initial allocation of local electricity, denoted by $A_{i}$, and expressed as:
\begin{gather}
	\label{eqn:initial_allocated_production}
	A_{i} = K_{i} \cdot \sum_{i \in \mathcal{I}} P^n_{t,i} \quad \forall (t,i) \in \mathcal{T} \times \mathcal{I}.
\end{gather}
Moreover, the tolerance around $K_{i}$ beyond which the optimal set of keys $k_{t,i}$ cannot deviate, is given by $X_{t,i}$:
\begin{gather}
	\label{eqn:tolerance_initial_optimised_keys}
	X_{t,i} = \left| k_{t,i} - K_{i} \right| \quad \forall (t,i) \in \mathcal{T} \times \mathcal{I}. 
\end{gather}

The initial set of keys and tolerance and, by extension, the initial allocation of local electricity generation can be integrated into Linear~program~\hyperref[lp:lp1]{1}, thus creating a new linear program.

\begin{algorithm}
\label{lp:lp3}
\caption{Computation of repartition keys with an initial set of keys.}

\begin{equation}
	\label{eqn:objective_function3}
	\displaystyle\min_{z \in Z}
	\sum_{i \in \mathcal{I}} \sum_{t \in \mathcal{T}} B_{t, i} + \xi^{d} \cdot \left( a_{t}^{+} + a_{t}^{-} \right),
\end{equation}
subject to:
\begin{gather}
	\nonumber \text{Equations \eqref{eqn:keys_computation} -- \eqref{eqn:limit_vars}}, \\
	\label{eqn:allocation_positive_deviation}
	a_{t,i} - A_{i} \leq a_{t}^{+} \quad \forall (t,i) \in \mathcal{T} \times \mathcal{I}, \\
	\label{eqn:allocation_negative_deviation}
 	A_{i} - a_{t,i} \leq a_{t}^{-} \quad \forall (t,i) \in \mathcal{T} \times \mathcal{I}, \\
	\label{eqn:deviation_keys}
	k_{t,i} - K_{i} \leq X_{t,i} \quad \forall (t,i) \in \mathcal{T} \times \mathcal{I}, \\ 
	\label{eqn:deviation_keys2}
	K_{i} - k_{t,i} \leq X_{t,i} \quad \forall (t,i) \in \mathcal{T} \times \mathcal{I}, \\
	a_{t}^{+}, a_{t}^{-} \in \mathbb{R}_{+}.
\end{gather}

\end{algorithm}

In this linear program, the vector of decision variables is $z = \left( k_{t,i}, y_{t,i}, a_{t,i}, v_{t,i}, a_{t}^{+}, a_{t}^{-} \right) \in Z$.

The objective function of this linear program (Equation~\eqref{eqn:objective_function3}) has as an additional term compared to Equation~\eqref{eqn:objective_function}.
This extra term is introduced to deal with cases with multiple solutions to the optimisation problem, which may occur, for instance, if the sum of the net consumption of the members of the REC is greater than the sum of the net production, and all members buy and sell energy at the same price to their electricity suppliers.
In such a context, this extra term favours a solution that distributes the local production equally among the REC members, something we believe is desirable.
Without this term, the allocation in these cases would be uneven, favouring some users depending on the optimisation solver numerical preferences.
The price $\xi^{d}$ is, in effect, fictive, and must be low (a fraction of the smallest among the rest of the price signals) to avoid interfering with the computation of the optimal repartition keys based on a minimisation of the electricity bill.

\section{Case study}
\label{sec:case_study}
This section introduces four different examples to study the performance of the linear programs presented in Sections~\ref{sec:rec_and_repartition_keys}~and~\ref{sec:rec_sustainability}.
Additionally, a complexity analysis has been carried out, and can be found in Appendix~\ref{app:complexity_analysis}.

The first two examples focus on Linear~program~\hyperref[lp:lp1]{1}---these cases illustrate the proposed methodology in (i) a simple numerical example to showcase its functioning, and (ii) a real test case showing a cost analysis derived from the use of this optimisation problem.
The third example analyses the introduction of revenue sharing through the computation of the SSR, following Linear~program~\hyperref[lp:lp2]{2}.
Finally, the fourth example focuses on Linear~program~\hyperref[lp:lp3]{3}, studying the impact of introducing initial contractual repartition keys on the costs of the REC members.
Whilst the first two examples showcase the general functioning of the proposed optimisation problem using repartition keys, the last two aim to illustrate the cost sharing mechanism behind the two alternative algorithms presented in this work.

To perform the simulations, a set of price signals is needed.
For simplicity, these price signals are the same for all the examples run in this paper.
Moreover, we assume that all REC members have similar retailer contracts both for imports and exports from and to the grid.
This does not imply that the ECM is the electricity supplier of the REC and, in fact, REC members have their own suppliers.
Table~\ref{tab:price_signals} lists these prices.

\begin{table}[!htb]
	\centering
	\caption{Price signals in EUR/MWh.}
	\begin{threeparttable}
		\begin{tabular}{lllll}
			\toprule
			$\xi_{i}^{b}$ & $\xi_{i}^{s}$ & $\xi_{i}^{l-}$ & $\xi_{i}^{l+}$ & $\xi_{i}^{d}\tnote{*}$ \\
			\midrule
			220 & 60 & 100 & 98 & 1 \\
			\bottomrule
		\end{tabular}
		\begin{footnotesize}
			\begin{tablenotes}
				\item [*] This price is only relevant for the last example shown in this section.
			\end{tablenotes}
		\end{footnotesize}
	\end{threeparttable}
	\label{tab:price_signals}
\end{table}

\subsection{Test case 1: performance on a simplified example}
\label{subsec:test_case_1}
The first example provides a simplified test case to acquaint the reader with the salient properties of our approach.
This example features an REC with two consumers (User1 and User2), one producer (User3) and one prosumer (User4).
The optimisation horizon is two metering periods, the first one with more production than consumption, and the second one with more consumption than production. Table~\ref{tab:initial_state_ex1} presents the inputs used for this simulation: positive and negative values correspond to consumption and electricity generation respectively.
The units in this example are kWh.

\begin{table}[!htb]
	\centering
	\caption{Test case 1 -- inputs.}
	\label{tab:initial_state_ex1}
	\begin{tabular}{ccccc}
		\toprule
		\textbf{Metering period} & \textbf{{\color{red}User1}} & \textbf{{\color{red}User2}} & \textbf{{\color{green}User3}} & \textbf{{\color{orange}User4}} \\
		\midrule
		\midrule
		\multicolumn{5}{c}{Consumption} \\
		\midrule
		2017-03-01 00:00 & 0.17 & 0.21 & -0.50 & 0.08 \\
		2017-03-01 00:15 & 0.21 & 0.23 & -0.30 & -0.02 \\
		\midrule
		\bottomrule
	\end{tabular}
\end{table}

These inputs are used to optimise the repartition keys and the allocation of local electricity generation to the REC members.
The results of this optimisation are presented in Table~\ref{tab:results_ex1} which lists the optimal set of repartition keys, the allocated generation, and the sold electricity (both within the REC and to the grid).
The set of optimal repartition keys leads to an optimal allocation of the generation among the REC members by which any deficit of local electricity generation is supplied by the retailers, whereas any excess is sold to them.
The optimal allocation of electricity generation is sold and delivered either as \emph{local sales} to REC members, or as \emph{global sales} to the REC members' electricity suppliers.

\begin{table}[!htb]
	\centering
	\caption{Test case1 -- outputs.}
	\label{tab:results_ex1}
	\begin{tabular}{ccccc}
		\toprule
		\textbf{Metering period} & \textbf{{\color{red}User1}} & \textbf{{\color{red}User2}} & \textbf{{\color{green}User3}} & \textbf{{\color{orange}User4}} \\
		\midrule
		\midrule
		\multicolumn{5}{c}{A -- Optimised repartition keys} \\
		\midrule
		2017-03-01 00:00 & 0.39 & 0.45 & 0.00 & 0.16 \\
		2017-03-01 00:15 & 0.47 & 0.53 & 0.00 & 0.00 \\
		\midrule
		\midrule
		\multicolumn{5}{c}{B -- Optimised verified allocated production} \\
		\midrule
		2017-03-01 00:00 & 0.17 & 0.21 & 0.00 & 0.08 \\
		2017-03-01 00:15 & 0.15 & 0.17 & 0.00 & 0.00 \\
		\midrule
		\midrule
		\multicolumn{5}{c}{C -- Electricity generation sold locally to the REC} \\
		\midrule
		2017-03-01 00:00 & 0.00 & 0.00 & 0.46 & 0.00 \\
		2017-03-01 00:15 & 0.00 & 0.00 & 0.30 & 0.02 \\
		\midrule
		\midrule
		\multicolumn{5}{c}{D -- Electricity generation sold to the grid} \\
		\midrule
		2017-03-01 00:00 & 0.00 & 0.00 & 0.04 & 0.00 \\
		2017-03-01 00:15 & 0.00 & 0.00 & 0.00 & 0.00 \\
		\midrule
		\bottomrule
	\end{tabular}
\end{table}

In Table~\ref{tab:results_ex1} we can see that for the first metering period, local sales (sub-table C) amount to 0.46, which is the total demand of the system (see Table~\ref{tab:initial_state_ex1}).
The electricity generation surplus is 0.04, and is sold as global sales (sub-table D).
In the second metering period, local sales (sub-table C) are 0.32, which correspond to the total available electricity generation (see Table~\ref{tab:initial_state_ex1}).
Since, at metering period two, there is greater demand than electricity generation, there are no global sales.
The distribution between local and global sales responds to the different price signals and, in particular, to the different spreads between $\xi_{i}^{b} - \xi_{i}^{s}$ and $\xi_{i}^{l-} - \xi_{i}^{l+}$ (see Equation \eqref{eqn:billing_costs} for the definition of the price signals).
These spreads represent the final price per kWh that REC members pay to consume electricity from their electricity suppliers in the first case, and from the REC in the second one.
In this example, the first spread is 0.16 EUR/kWh and the second one is 0.002 EUR/kWh, consequently, the optimisation problem allocates as much as the local electricity generation as possible as local sales first, and then the rest as global ones.

\subsection{Test case 2: performance on a realistic example}
\label{subsec:test_case_2}
This second analysis introduces a more realistic set-up where an REC with 23 consumers and one producer is simulated over one year of operation.
Input consumption data corresponds to real measurements of small- and medium-volume electricity consumers in Belgium over one year (2017).
In Appendix~\ref{app:consumption_production_data}, the consumption and production data is illustrated in detail.

\begin{figure}[!htb]
	\centering
	\includegraphics[width=0.95\columnwidth]{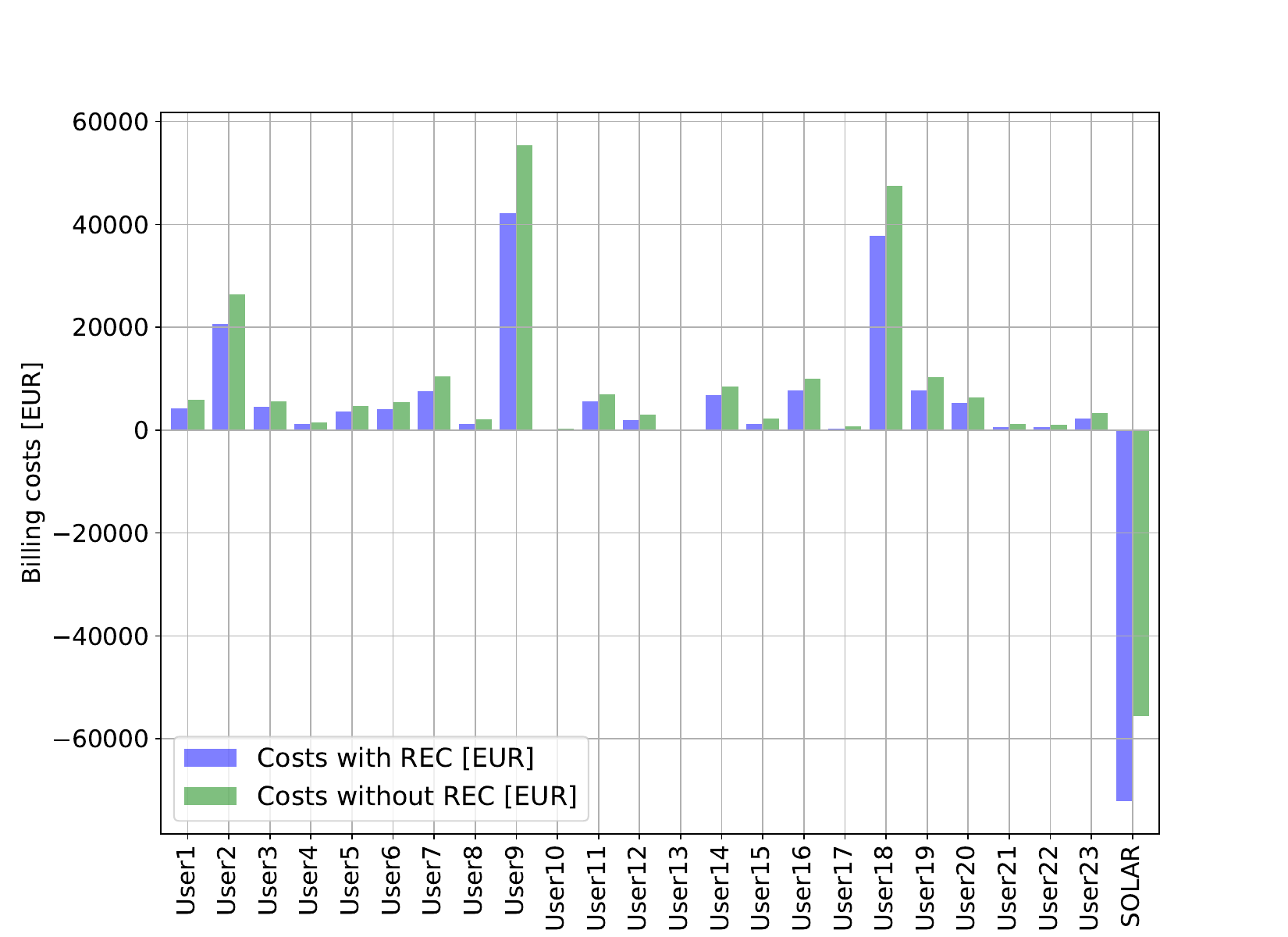}
	\caption{Annual electricity bills of the REC members. A positive value represents a cost, whilst a negative value represents a revenue. Note that in this particular example there are 23 consumers and one producer.}
	\label{fig:costs_final_customers}
\end{figure}

With these data, we have run the linear program presented in Section~\ref{subsec:problem_formulation} (Linear~program~\hyperref[lp:lp1]{1}) to optimise the repartition keys and, as such, the allocation of total available local electricity generation to each REC member.
With the optimal allocations of the REC members, we have computed the annual electricity bills of each of them, as per Equation~\ref{eqn:billing_costs}.
In addition, we have also computed the annual electricity bills of each user without REC, that is, in a situation where all of their consumption and generation is purchased or sold via their traditional retailer contracts, without considering any local electricity market.
These two electricity bills can then be compared and are reported in Figure~\ref{fig:costs_final_customers}.

In this figure, positive values imply a cost, whilst negative values imply a revenue for the REC members.
We can see that the deployment of an REC induces cost reductions in annual electricity bills for all REC members.
Furthermore, the revenue (negative costs) of the only producer (\texttt{SOLAR}) is increased when the REC is created.
In Table~\ref{tab:cost_analysis_test2}, a cost analysis is shown where the total electricity billing costs of the REC and the average billing costs and revenue of the REC members are shown.

\begin{table}[!htb]
	\centering
	\caption{Test case 2 -- Cost comparison between total and average electricity bills (accounting for sales) with and without REC. The first column corresponds to the aggregated bill, the second column correspond to the average bill for the consumers, and the third column corresponds to the average nevative cost (or revenue) for the producers.}
	\label{tab:cost_analysis_test2}
    \resizebox{\columnwidth}{!}{%
	\begin{tabular}{llll}
		\toprule
		\textbf{Costs [TEUR]} & \textbf{Total} & \textbf{Avg. consumers} & \textbf{Avg. producers} \\
		\midrule
		\textbf{w/o REC} & $163.67$ & $9.53$ & $-55.62$ \\
		\textbf{w/ REC} & $95.00$ \textcolor{mygreen}{($-42$\%)} & $7.27$ \textcolor{mygreen}{($-23.8$\%)} & $-72.13$ \textcolor{mygreen}{($+29.7$\%)} \\
		\bottomrule
	\end{tabular}%
	}
\end{table}

The results in Table~\ref{tab:cost_analysis_test2} show that creating an REC and determining the allocation of total available local production through an \emph{ex-post} optimisation of the repartition keys, may lead to significant overall savings for the REC.
In particular, in our test case, overall savings of up to 42\% are attained.
These overall savings take into account both the reduced costs for \texttt{User1} -- \texttt{User23} ($23.8$\%) as well as the increased revenue of producer \texttt{SOLAR} ($29.7$\%).
The results obtained through this analysis greatly depend on the price signals associated to both retailer contracts and internal prices (see Table~\ref{tab:price_signals}).
How to compute or select these prices, though, falls out of the scope of our work, which aims to show a methodology for allocating internal electricity flows based on a known set of prices.
For an analysis of the computation of these prices, the reader can study the following works \cite{long2017peer,moret2018energy,cornelusse2019community,grzanic2021electricity}---the set of assumptions needed in all these studies, and in particular the need for the ECM to act also as the single electricity supplier of all REC members, make them not compliant with the latest European regulations.
Since one of the main goals of out work is to produce a framework compliant with these regulations, in this paper we abstract from any computation of the internal prices and rather fix them in between the range given by an average retailer price for consumption (220 EUR/MWh) and a wholesale price for sales (60 EUR/MWh).

\subsection{Test case 3: minimum SSR}
\label{subsec:test_case_3}
After evaluating the performance of Linear~program~\hyperref[lp:lp1]{1}, the third example, introduced in this section, analyses how the revenue can be shared among the REC members through the computation of the SSR, as explained in Section~\ref{subsec:ssr}, following Linear~program~\hyperref[lp:lp2]{2}.
To that end, we make use of the same data set as in the previous test case (i.e., 23 consumers and one producer).

\begin{figure}[!htb] 
	\centering
	\subfloat[Without minimum SSR bound (i.e. $SSR_{i}^{min = 0\%}$).]{{\includegraphics[width=0.95\columnwidth]{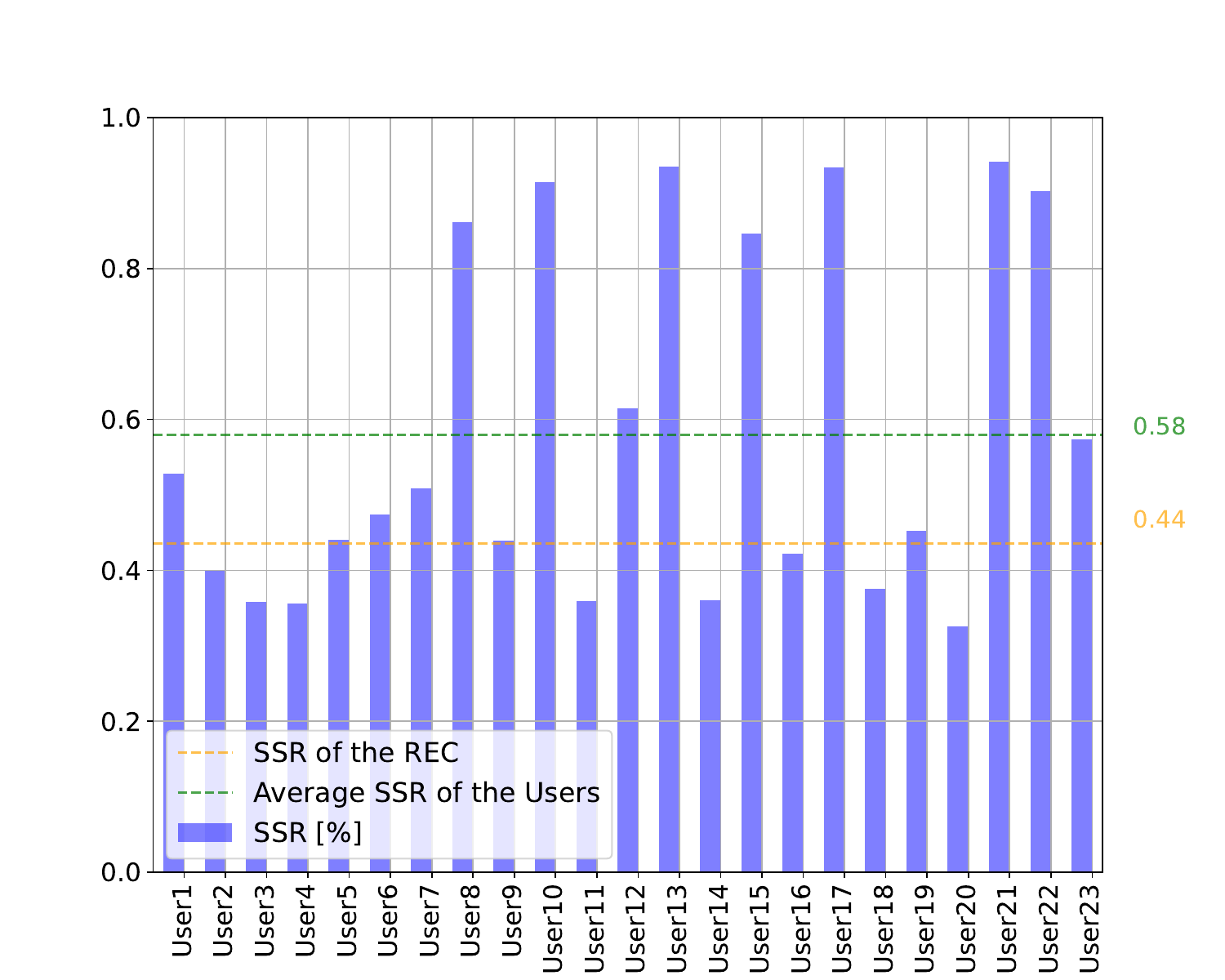}}
	\label{subfig:min_ssra}
	} \quad
	\subfloat[With an enforced minimum SSR of $42\%$ (i.e. $SSR_{i}^{min = 42\%}$)]{{\includegraphics[width=0.95\columnwidth]{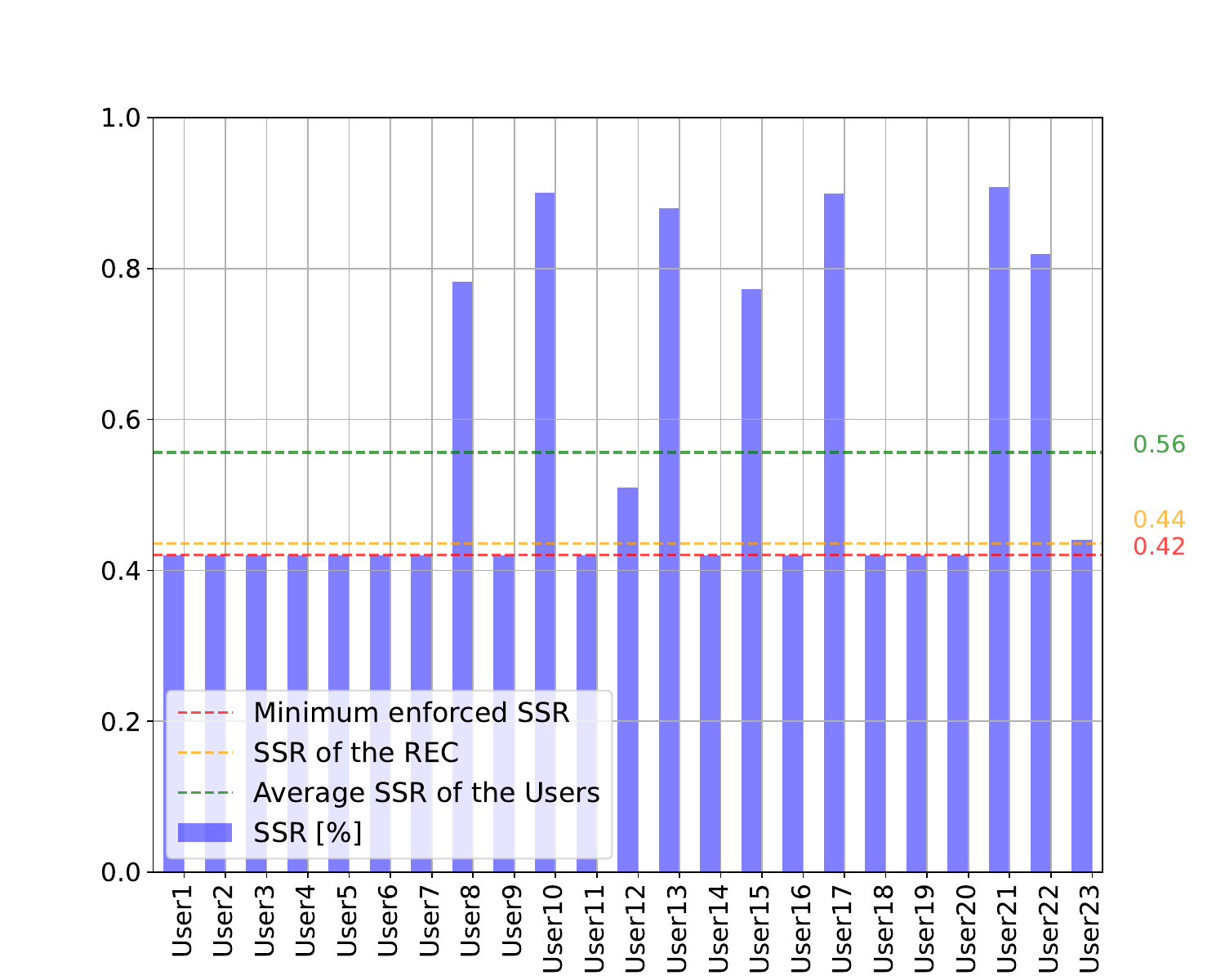}}
	\label{subfig:min_ssrb}
	}
	\caption{SSR of the REC members before and after enforcing a minimum SSR for all of them of 42\%.}
	\label{fig:min_ssr}
\end{figure}

To analyse the revenue sharing mechanism through the computation of the SSR we have first solved Linear~program~\hyperref[lp:lp2]{2}, without enforcing the SSR constraint (Equation~\eqref{eqn:self_sufficiency_rate}).
This means that at first this constraint is not tight and the SSR of the REC members is freely optimised to minimise the overall REC billing costs.
Then, we have enforced the bound by selecting positive values of $SSR_{i}^{min}$, effectively tightening Equation~\eqref{eqn:self_sufficiency_rate} for all REC members at the same time.
The values can be then progressively increased until reaching the maximum possible SSR before inducing infeasibilities. 
In our example, this occurs at an SSR of $42$\%, and any value on the left hand side of Equation~\eqref{eqn:self_sufficiency_rate} greater than this leads to a non feasible space of solutions.
The findings of this analysis are presented in Figure~\ref{fig:min_ssr}.

\paragraph*{No minimum SSR (Figure~\ref{subfig:min_ssra}):}
In this figure we observe that the values of $ssr_{i}$ span from $32.6$\% for \texttt{User20} to $94.1$\% for \texttt{User21}.
In terms of billing costs, this means that $32.6$\% of the electricity consumption of \texttt{User20} will be paid at internal price (i.e., $\xi_{i}^{l-}$) and $67.4$\% at retail price (i.e., $\xi_{i}^{b}$).
On the other hand, \texttt{User21} will pay $94.1$\% of the electricity consumption at internal price, and $5.9$\% at retail price.
The final electricity bill of the REC members will depend on the spread between retail and internal prices for each REC member.
A positive spread (retail minus internal) may induce bill savings for the REC members when comparing to the case without REC.
In our example, if no minimum SSR is enforced, the creation of an REC induces bill savings of $17.8$\% to \texttt{User20} and $51.3$\% to \texttt{User21}.
These savings are computed with respect to the traditional retailer contracts the members would have without an REC.

\paragraph*{Minimum SSR of $42$\% (Figure~\ref{subfig:min_ssrb}):}
In this figure, the values of $ssr_{i}$ span from $42$\% (\texttt{User1} -- \texttt{User7}, \texttt{User9}, \texttt{User11}, \texttt{User14}, \texttt{User16}, and \texttt{User18} -- \texttt{User20}) to $90.1$\% (\texttt{User21}).
Following a similar analysis as the previous one, and focusing on the same two users (\texttt{User20} and \texttt{User21}), we can see that creating an REC and enforcing a minimum SSR of $42$\% to all REC members induces savings of $22.9$\% to \texttt{User20} and $49.6$\% to \texttt{User21}.
This indicates that enforcing a minimum SSR has a redistributional effect of the bill savings among the REC members.
In particular, \texttt{User20} sees the bill savings increase by $5.1$\% compared to the case where no SSR is enforced whereas \texttt{User21} sees the bill savings decrease by $1.7$\%.

\begin{figure}[!htb]
	\centering
	\includegraphics[width=0.95\columnwidth]{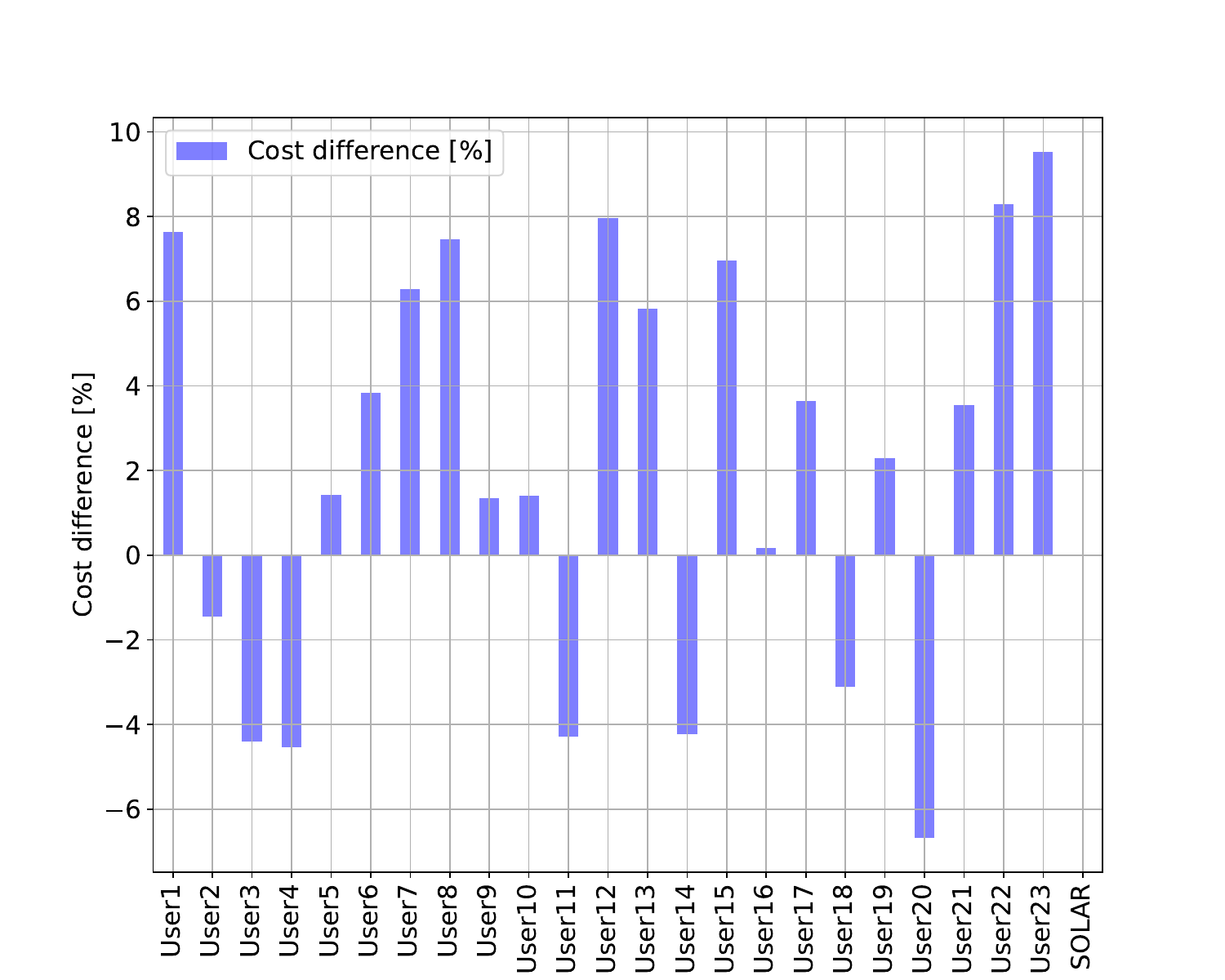}
	\caption{Difference in the billing costs of REC members, with and without enforcing any minimum SSR ($42$\%).}
	\label{fig:costs_ssr}
\end{figure}

An important remark of the redistribution of local electricity generation taking place as a result of enforcing a minimum SSR is that, since the total available local electricity remains constant (these optimisations are performed \emph{ex-post} with real consumption and generation data which are fixed and not optimised), enforcing a minimum SSR can be seen as a zero-sum game where in order to increase the SSR of some REC members, it must be withdrawn from others.
This implies that enforcing a high minimum SSR for all REC members at once results in a flattened distribution of the SSR across the REC members, as can be observed moving from Figure~\ref{subfig:min_ssra} to Figure~\ref{subfig:min_ssrb}.
The consequence of this flattening is that, even though the SSR of the REC is constant ($44$\%) as it depends solely on the total available local electricity, the average SSR of the REC members is eroded, decreasing from $58$\% to $56$\%.
This decrease has no impact on the overall billing costs of the REC, which depend on the SSR of the REC (constant), however, it affects how these costs are distributed, creating a divergence with respect to the case where no minimum SSR is enforced.
This divergence can be quantified, and is illustrated in Figure~\ref{fig:costs_ssr} which showcases the difference between the billing costs of all REC members when $SSR_{i}^{min} = 0$\% and when $SSR_{i}^{min} = 42$\%.

Figure~\ref{fig:costs_ssr} shows that members who are forced to give up part of their SSR ($ssr_{i}$) when enforcing a minimum SSR ($SSR_{i}^{min}$), incur higher costs than before enforcing it.
Conversely, those members who see their ssr increased after enforcing a minimum SSR, incur lower costs than before.
In this example, the REC members' gains range from $0.25\%$ for \texttt{User16} to $9.5\%$ for \texttt{User23}, whereas the losses range from $-1.5\%$ for \texttt{User2} to $-6.5\%$ for \texttt{User20}.
In this figure, we can also observe that the difference between total gains and total losses is asymmetrical, the losses (positive values) outweighing the gains (negative values)\footnote{These values represent a fictitious cost difference for the same REC when applying different minimum SSR rules and are not comparable with the costs reported when explaining Figure~\ref{fig:min_ssr}.}, which is consistent with the decrease in average SSR from $58$\% to $56$\%.

\subsection{Test case 4: impact of initial repartition keys}
\label{subsec:test_case_4}
In this last example we analyse the minimum revenue of REC members obtained through the use of initial repartition keys as described in Section~\ref{subsec:initial_keys}, Linear~program~\hyperref[lp:lp3]{3}.
To that end, we must define the initial keys $K_{i}$ and the tolerance $X_{t,i}$.

In this example, we test two different types of initial repartition keys ($K_{i}$):
\begin{itemize}
	\item Uniform: evenly distributed among the REC members---all members with positive net demand receive the same percentage of the total available local electricity.
	\item Proportional: each member obtains a percentage of the total local available electricity which is proportional to their average demand over the simulated period---each member receives a different initial key, constant over time.
\end{itemize}
Both sets of initial keys can be pre-computed by the ECM based on historical data of the REC members.

As for the tolerance ($X_{t,i}$), in this section's analyses, we have run a sensitivity to several values: $\left[ 0, 1, 3, 5, 10, 20, 30, 50, 100 \right]$.
All these values are given in percentages of the initial repartition keys---they indicate how much (in percentage) the optimal set of keys can deviate from the given initial keys (from $0$\% to $100$\%).
A small tolerance constraints the space of solutions of the optimal set of keys whereas a large tolerance provides more freedom to the optimisation.

With these parameters, we have analysed three effects on the REC: (i) the level of use of the internal market (i.e., amount of exchanges within the REC relative to the total exchanges); (ii) the billing costs for the REC members, and (iii) the allocated local electricity among the REC members.
To carry out these analyses, a small REC is selected, composed of six REC members: five consumers (\texttt{User1} -- \texttt{User5}) and one producer (\texttt{User6}).
The simulation horizon is one month (April 2017).
More information concerning the simulation data can be found in Appendix~\ref{app:consumption_production_data}.

\subsubsection{Level of use of the internal market}
\begin{figure}[!htb]
	\centering
	\includegraphics[width=0.95\columnwidth]{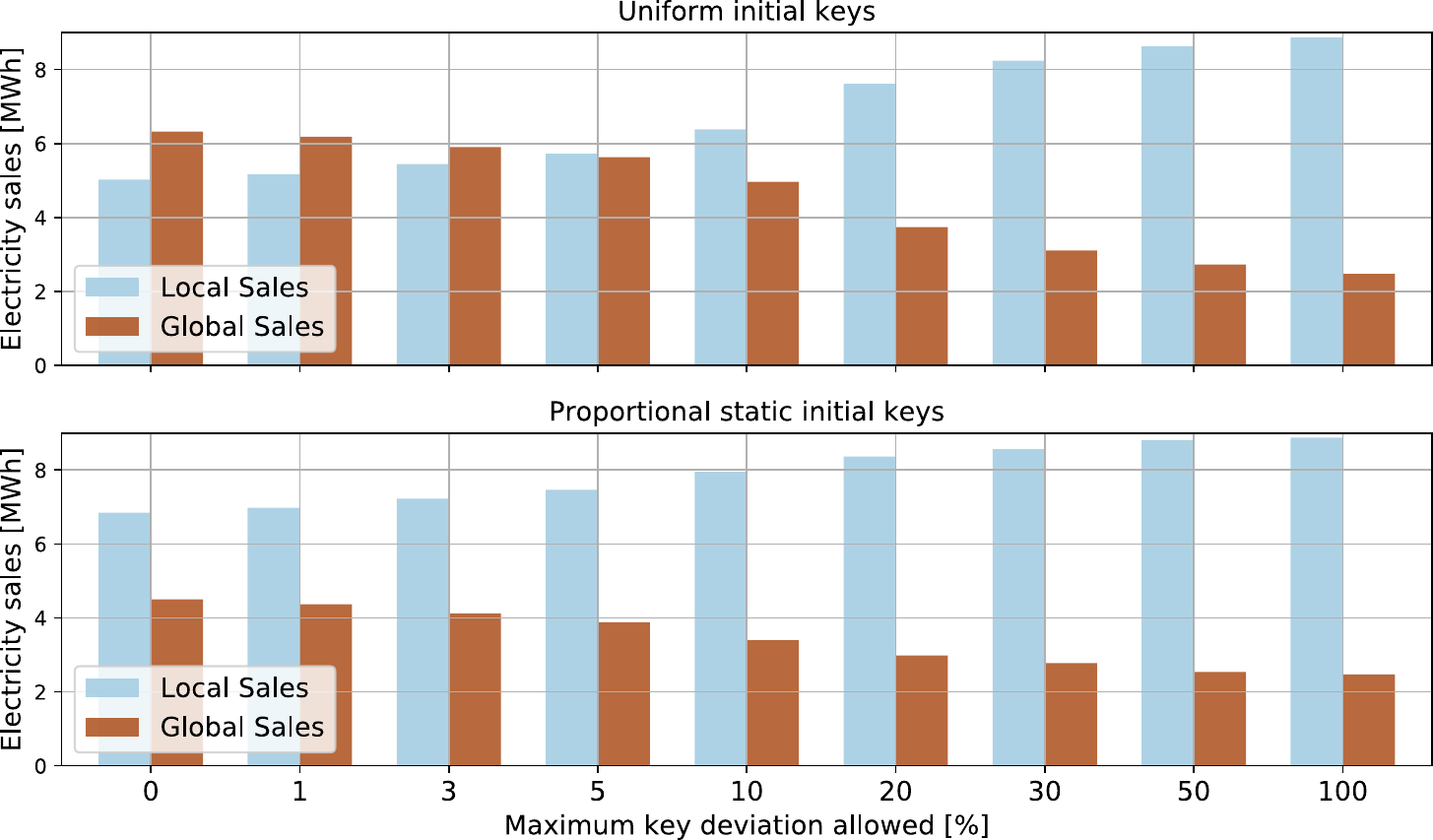}
	\caption{Total electricity sold in the internal REC market (local sales) and to the retailer (global sales) for a range of tolerances $X_{t,i}$.}
	\label{fig:sales}
\end{figure}

To evaluate the level of use of the internal market, we have computed the spread between electricity sales in the internal market, or local sales ($y_{t,i}$) and electricity sales to the retailer, or global sales ($P_{t,i}^{n} - y_{t,i}$).
This spread is strictly positive if local are greater than global sales.
This analysis is presented in Figure~\ref{fig:sales}

Figure~\ref{fig:sales} shows that for the two sets of initial repartition keys tested (uniform and proportional), setting a maximum tolerance $X_{t,i}=100$\% meaning that the optimised keys are free to change, leads to the same spread between local and global sales (right hand side of Figure~\ref{fig:sales}).
However, they differ when $X_{t,i}<100$\%.
When applying proportional static initial keys, the spread is always positive and increases with the value of $X_{t,i}$.
On the other hand, applying uniform initial keys leads to negative spreads when $X_{t,i}<5$\%.
This analysis shows that when no limitation on the tolerance $X_{t,i}$ is imposed, the optimisation problem finds the same solution regardless of the initial keys.
However, when this constraint is tight ($X_{t,i}<100$\%), the selection of initial keys has a notable impact on the results.
From the REC members perspective, proportional initial repartition keys favour local sales more than uniform ones---assuming that the local market is economically beneficial with respect to the grid, proportional initial keys are desirable as they lead to lower costs for the REC members.

\subsubsection{Billing costs of the REC members}
\begin{figure}[!htb]
	\centering
	\includegraphics[width=0.95\columnwidth]{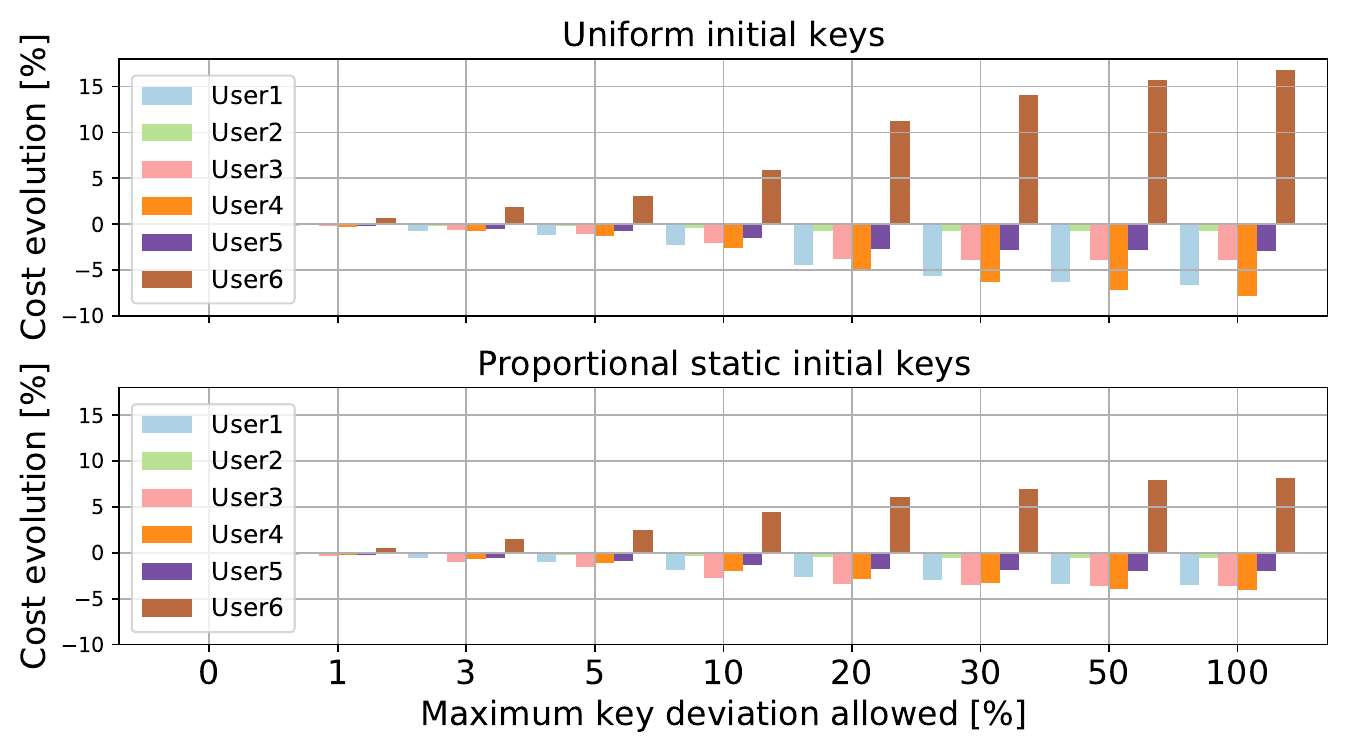}
	\caption{Costs of the REC members for a range of tolerances $X_{t,i}$, relative to the costs when $X_{t,i}=0$.}
	\label{fig:costs_deviations}
\end{figure}

In this analysis we evaluate the individual costs of the REC members using the range of tolerances $X_{t,i}$ and the two types of initial keys previously exposed.
Results of this analysis are shown in Figure~\ref{fig:costs_deviations}.

In this Figure we can see the billing cost evolution of all REC members for an gradually increasing value of the tolerance ($X_{t,i}$).
This evolution is shown in relative increase (or decrease) compared to the billing cost when the tolerance is $0$\% (i.e., when the initial and the optimised keys are the same).
In Figure~\ref{fig:costs_deviations}, positive values correspond to revenue, while negative values correspond to costs.
In this example we can observe that, as the tolerance constraint is loosened (increasing the value of $X_{t,i}$), the costs decrease for \texttt{User1} -- \texttt{User5} decrease and the revenue increases for \texttt{User6}.
Moreover, these trends are similar regardless of the initial repartition keys used, however, the magnitude of costs and revenue are different, and are significantly larger when using uniform instead of proportional initial repartition keys.
The savings of \texttt{User1} -- \texttt{User5} for uniform keys span from $1$\% to $8$\%, whereas for proportional they span from $0.5$\% to $4$\%.
Likewise, the increase in revenue of \texttt{User6} is up to $16\%$ with uniform keys and $8\%$ with proportional keys.
The reason for these significant differences lies on the initial solution provided by the two types of initial repartition keys---uniform keys provide an initially worse solution (in terms of costs and revenue) than proportional keys, therefore when the tolerance is increased (effectively allowing more freedom to the optimisation program to compute the optimised keys), the improvement in costs and revenue is more apparent when uniform initial keys are employed.
This effect reinforces our previous conclusion that using proportional initial keys, that is, keys that are proportional to the demand of the REC members, seems to be a good practice, which concurs with current practices as seen in \cite{frieden2019overview}.
This remark holds true as long as the billing costs are based mainly on volumetric charges which depend on energy consumption, for other types of charges (capacity, fixed, or time-of-use), more analyses are needed.

\subsubsection{Allocation of local electricity}
\begin{figure}[!htb]
	\centering
	\includegraphics[width=\columnwidth]{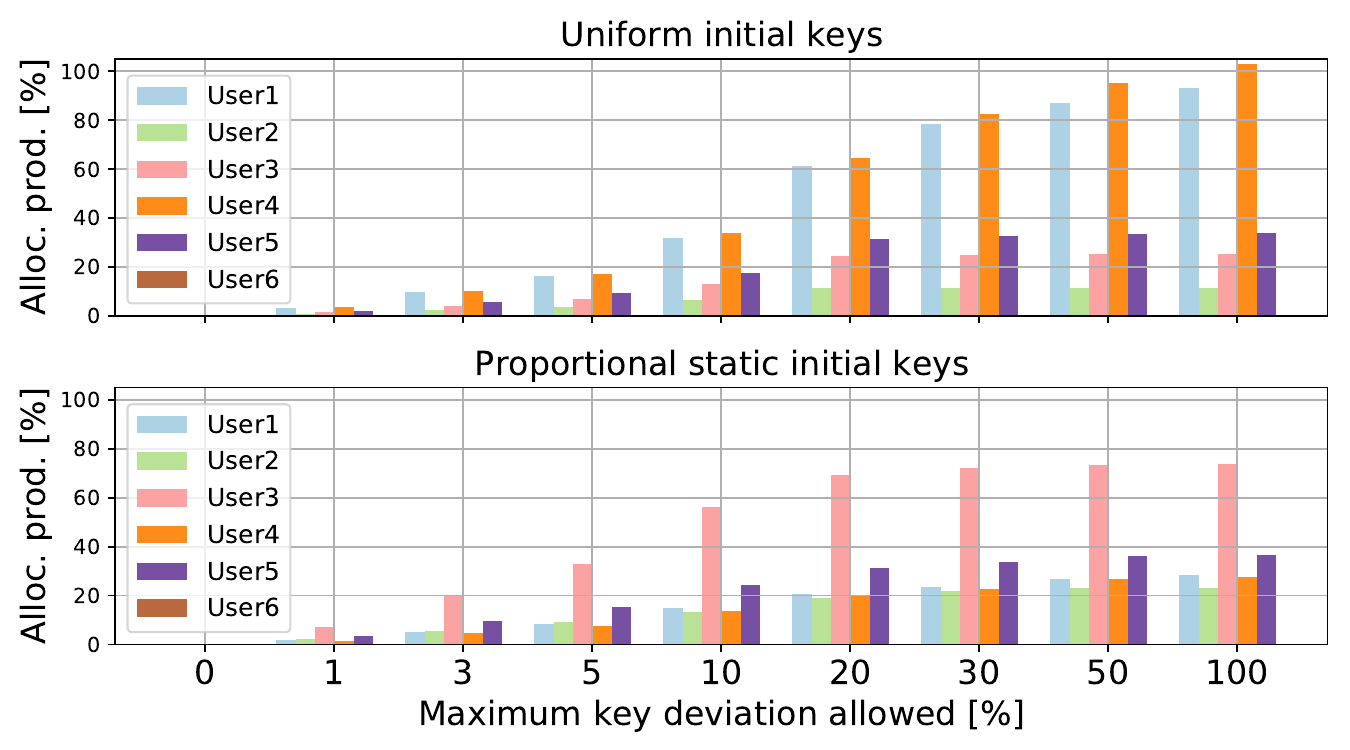}
	\caption{Allocated local electricity generation of the REC members for a range of tolerances $X_{t,i}$ relative to the allocated electricity when $X_{t,i}=0$.}
	\label{fig:verified_production}
\end{figure}

In this analysis we show the allocation of local electricity (for the same range of tolerances $X_{t,i}$ as in the previous analyses) as its relative increase (or decrease) with respect to the allocation when the tolerance is $0$\% (i.e., when initial and optimised keys are equal).
This analysis is presented in Figure~\ref{fig:verified_production}.

The trends followed by the REC members' allocated production are similar for both uniform and proportional keys---in both cases this trend is upward when the value of the tolerance $X_{t,i}$ is gradually relaxed (i.e., increased from $0$\% to $100$\%).
However, the magnitudes of these trends and the members involved are different---whilst with uniform initial keys the allocated electricity reaches around $100$\% for \texttt{User1} and \texttt{User4}, with proportional ones the allocated electricity only reaches around $70$\% for \texttt{User3}.
The difference in these results stems from the different consumption profiles of the REC members and the way the initial keys allocate the initial local electricity generation.
Uniform keys allocate the same local electricity to all REC members, whereas proportional keys allocate the electricity based on the average demand of each REC member with respect to the overall demand of the REC.
In this case, the average demand of \texttt{User3} is lower than the other members, thereby the proportional keys provide a low initial electricity allocation.
This low initial allocation might not be the most cost-effective solution of the problem though.
For instance, the consumption of \texttt{User3}, although lower than the rest, is better synchronised with the generation and, consequently, relaxing the tolerance (increasing it), provides more room for the optimisation problem to find a better (more economic) solution by increasing the allocated local electricity to this member.
On the other hand, when using uniform keys, the allocated local electricity is not enough to cover the demand of \texttt{User1} and \texttt{User4}, and when the constraint is relaxed more allocation is provided to these two members.
Note that regardless of the initial keys, relaxing the constraint on the tolerance leads to a better (less costly) allocation of local electricity for all REC members, matching more efficiently supply and demand.
However, the initial keys play an important role ensuring a minimum allocation (and therefore a minimum revenue), potentially fostering the rollout of RECs.
Moreover, the magnitude of the changes in allocated local electricity when the tolerance is increased with uniform keys is larger than with proportional ones, once again suggesting that the later lead to a better initial solution, as other authors (see for instance \cite{frieden2019overview}) have already highlighted.

\section{Conclusion}\label{sec:conclusion}

Renewable energy communities (RECs) as a form of local electricity market have gained acceptance over the last years, e.g., in Europe new legislation is being developed to create the necessary sets of rules framing RECs' functioning.
In this context, this paper proposes a flexible modelling solution to the financial optimisation of RECs that complies with current regulations and can readily be adopted by REC managers.
This solution relies on the use of \emph{repartition keys}, representing proportions of overall local electricity generation available within the REC.
In our methodology, the surplus of electricity of REC members is aggregated and allocated to the rest of the REC members through the computation of repartition keys, thus optimally distributing the available local electricity among them.
The keys are optimally computed through an optimisation in the form of a linear program whose main decision variables are the repartition keys, and that takes place after physical delivery of electricity (using real data of electricity consumption and generation of the REC members).
Then, the billing process of the REC takes them into account to send the correct bills to each REC member, thus effectively optimising their financial flows \emph{ex-post}, which, for each REC member, are broken into two: (i) the lump sum owed to the REC, and (ii) the rest of the bill corresponding to traditional retailing contracts.
The repartition keys, therefore, determine the optimal trade-off between these two flows for each REC member.
Furthermore, to enhance the economic sustainability of the REC, this paper proposes two additions to this algorithm to (i) share the revenue among the REC members, and (ii) ensure a minimum revenue for each of them.
The revenue sharing is achieved through the computation of self-sufficiency rates, which indicate the amount of electricity demand covered by the REC with respect to the total demand.
Ensuring a minimum revenue is done through initial repartition keys used as a base for the computation of the optimised ones.
Thanks to these two approaches, the REC members can estimate the economic benefits they will derive from the REC, prior to their participation, thus enhancing the probability of a stable community.

Various test cases illustrate and test this methodology. The main findings of these tests are:
\begin{itemize}
	\item the use of repartition keys to optimise the financial flows of RECs is a practical approach to model RECs, and can be readily adopted by REC managers;
	\item establishing an REC can be economically beneficial for all REC members, provided that the REC internal exchanges are priced in between the retail purchasing and selling prices;
	\item the stability of an REC can be improved by making use of revenue sharing or minimum revenue methodologies (two examples of which are exposed in this work) which can enhance the economic benefits some of them derive from the REC;
	\item when using initial repartition keys, our experiments suggest that allocating the local electricity proportionally to the average historical demand of REC members is a good approximation, especially when the retail tariffs of the members are similar, which concurs with current practices in the literature.
\end{itemize}

The methodology presented in this paper has been tested and is currently being implemented by industrial partners in different REC managed by them.
After discussing the \emph{ex-post} optimisation of financial flows, a more comprehensive approach where the control of physical flows within the REC is also accounted for is a potential way to expand our work.

\appendices
\section{Complexity analysis}
\label{app:complexity_analysis}

In the final section of the results, we present an analysis of the complexity of the methodology proposed.
The number of constraints of the optimisation is $N_{cons}=9 |\mathcal{T}| |\mathcal{U}| + |\mathcal{T}| + |\mathcal{U}|$ and the number of variables $N_{var}=17 |\mathcal{T}| |\mathcal{U}| + 2 |\mathcal{T}| + |\mathcal{U}|$. Table~\ref{tab:running_time} introduces the running times for different complexities, ranging from 15 days with 10 REC members to one month with 100 members.
The optimisation problem is implemented with Pyomo in Python 3.8 and solved with the open source solver CBC.
Simulations are performed on a GNU/Linux machine with an Intel® Core™ i7-8665U and 16 Gb of RAM.

\begin{table}[!htb]
	\centering
	\caption{Running times of the proposed algorithm.}
	\resizebox{\columnwidth}{!}{%
	\begin{tabular}{rrrrrr}
		\toprule
		$|\mathcal{T}|$ & $|\mathcal{U}|$ &  $N_{cons}$ & $N_{var}$ & Build time [s] & Solve time [s] \\
		\midrule
		1,440 & 10 & 131,050 & 247,690 & 5.01 & 5.96 \\
		2,880 & 10 & 262,090 & 495,370 & 9.71 & 12.23 \\
		1,440 & 50 & 649,490 & 1,226,930 & 20.36 & 27.72 \\
		2,880 & 50 & 1,298,930 & 2,453,810 & 43.55 & 56.55 \\
		1,440 & 100 & 1,297,540 & 2,450,980 & 39.67 & 58.93 \\
		2,880 & 100 & 2,594,980 & 4,901,860 & 85.92 & 133.93 \\
		\bottomrule
	\end{tabular}
	}
	\label{tab:running_time}
\end{table}

\section{Linearisation of the SSR}
\label{app:linearisation_ssr}
Focusing on the numerator in Equations~\eqref{eqn:self_sufficiency_rate_def}~and~\eqref{eqn:self_sufficiency_rate}:

If $P_{t,i} > C_{t,i}$, the net consumption is $C^n_{t,i} = 0$ and thereby $v_{t,i} = 0$ as per Equation~\eqref{eqn:verified_allocation_limit2}.
In this case, the two numerators become equal to $\min \left\{ P_{t,i}, C_{t,i} \right\}$.

If $P_{t,i} \leq C_{t,i}$, the net consumption is $C^n_{t,i} \geq 0$, and thereby $v_{t,i} \geq 0$.
Then, by definition of $v_{t,i}$:
\begin{align}
	v_{t,i} & \leq C^n_{t,i} = C_{t,i} - P_{t,i} \\
	P_{t,i} + v_{t,i} & \leq P_{t,i} + C^n_{t,i} = C_{t,i}.
\end{align}
As $P_{t,i} + v_{t,i} \leq C_{t,i}$, the numerator in Equation \eqref{eqn:self_sufficiency_rate_def} becomes:
\begin{gather}
	\min \left\{ P_{t,i} + v_{t,i}, C_{t,i} \right\} = P_{t,i} + v_{t,i},
\end{gather}
which is equal to $\min \left\{ P_{t,i}, C_{t,i} \right\} + v_{t,i}$ since $P_{t,i} \leq C_{t,i}$.

\section{Consumers data}
\label{app:consumption_production_data}
\begin{figure}[!htb]
	\centering
	\includegraphics[width=0.99\columnwidth]{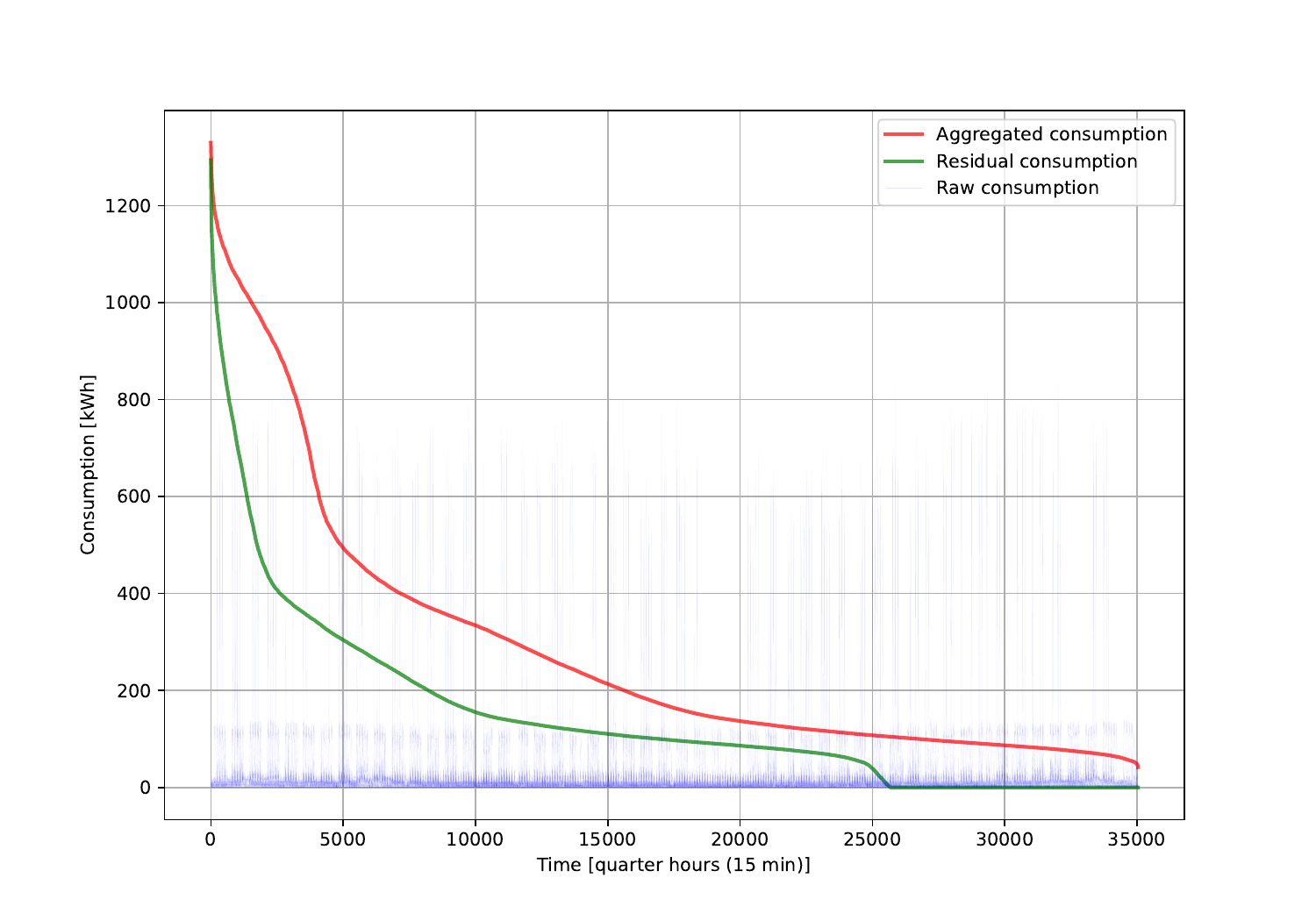}
	\caption{Consumption profiles of the REC members (blue), aggregated consumption of all REC members in a load duration curve (red), and residual consumption in a load duration curve (green). One year data: used in test cases \ref{subsec:test_case_2} and \ref{subsec:test_case_3}.}
	\label{fig:ldc}
\end{figure}

\begin{figure}[!htb]
	\centering
	\includegraphics[width=0.99\columnwidth]{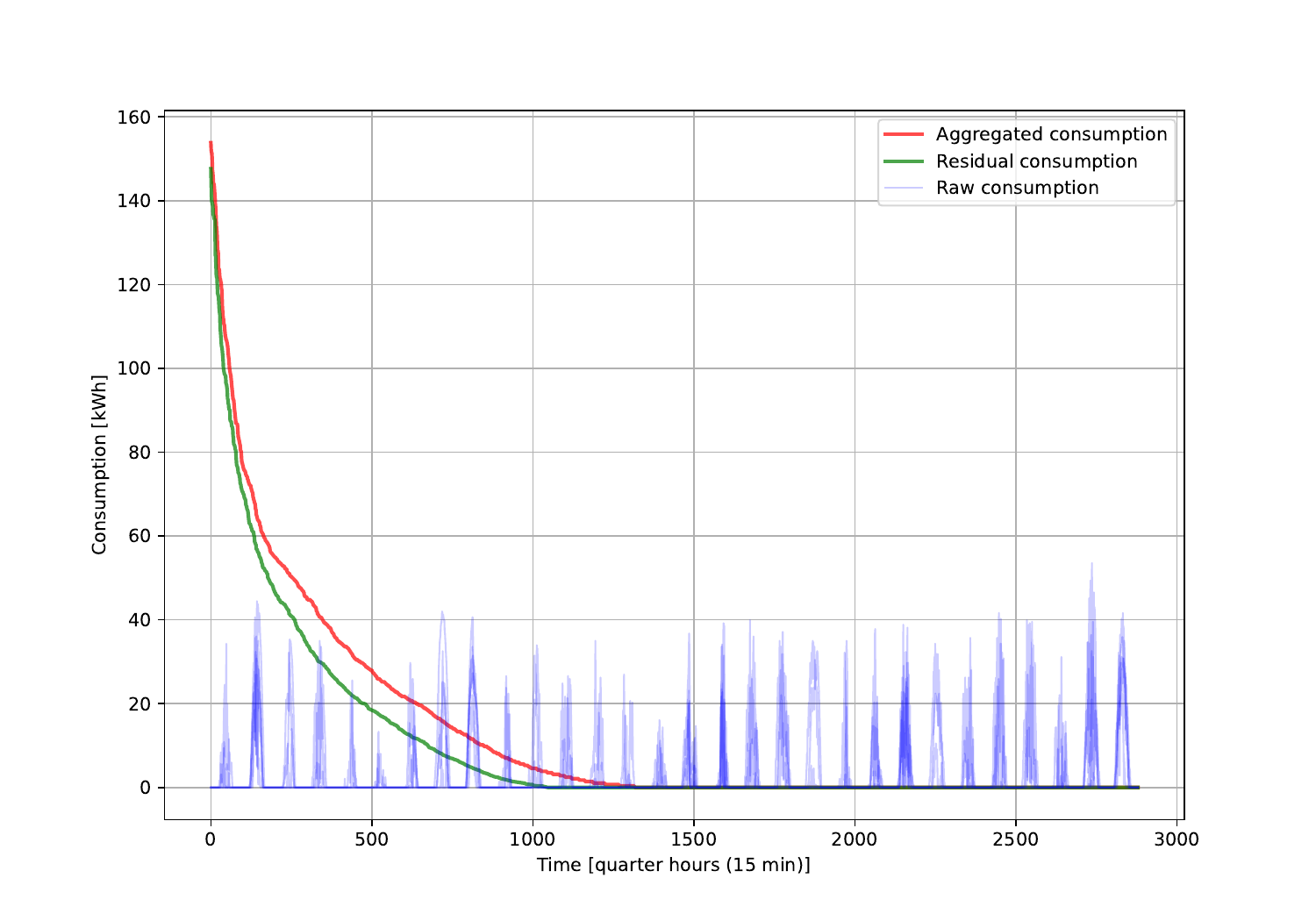}
	\caption{Consumption profiles of the REC members (blue), aggregated consumption of all REC members in a load duration curve (red), and residual consumption in a load duration curve (green). One month data: used in test case \ref{subsec:test_case_4}.}
	\label{fig:ldc_2}
\end{figure}

In Figures~\ref{fig:ldc}~and~\ref{fig:ldc_2}, the consumption and production data of the REC members for the one year and one month analyses respectively, is summarised.
In these figures, the raw data of electricity consumption of the REC members is presented in blue.
In addition, two load duration curves (LDC) are shown: the LDC of aggregated electricity consumption of the REC members can be seen in red, and the LDC of aggregated electricity consumption subtracting aggregated electricity generation, that is, the residual consumption of the system, is shown in green.

Figure~\ref{fig:ldc} presents the data corresponding to test cases \ref{subsec:test_case_2} and \ref{subsec:test_case_3}.
In this figure we can observe that the solar generation covers the demand of the REC roughly 60 days of the year.
The rest of the time, the presence of solar generation reduces the residual consumption significantly.
In particular, we can see that the peak consumption (left of the curves) is significantly reduced with solar generation, up to 40-50\%, whereas the baseload is only reduced by 10--20\%.

Figure~\ref{fig:ldc_2} presents the data corresponding to test case \ref{subsec:test_case_4}.
In this figure we can observe that the solar generation covers the demand of the REC roughly three days of the month.
The rest of the time, the presence of solar generation reduces the residual consumption evenly, slightly more on the baseload ($\approx 0$\%) than on the peaks ($\approx 5$\%).
In this case, the baseload does not cover the whole period, since the electricity load of the five consumers is or negligible for about half of the month (small consumers with little to no consumption during the night).

\begin{IEEEbiography}[{\includegraphics[width=1in,height=1.25in,clip,keepaspectratio]{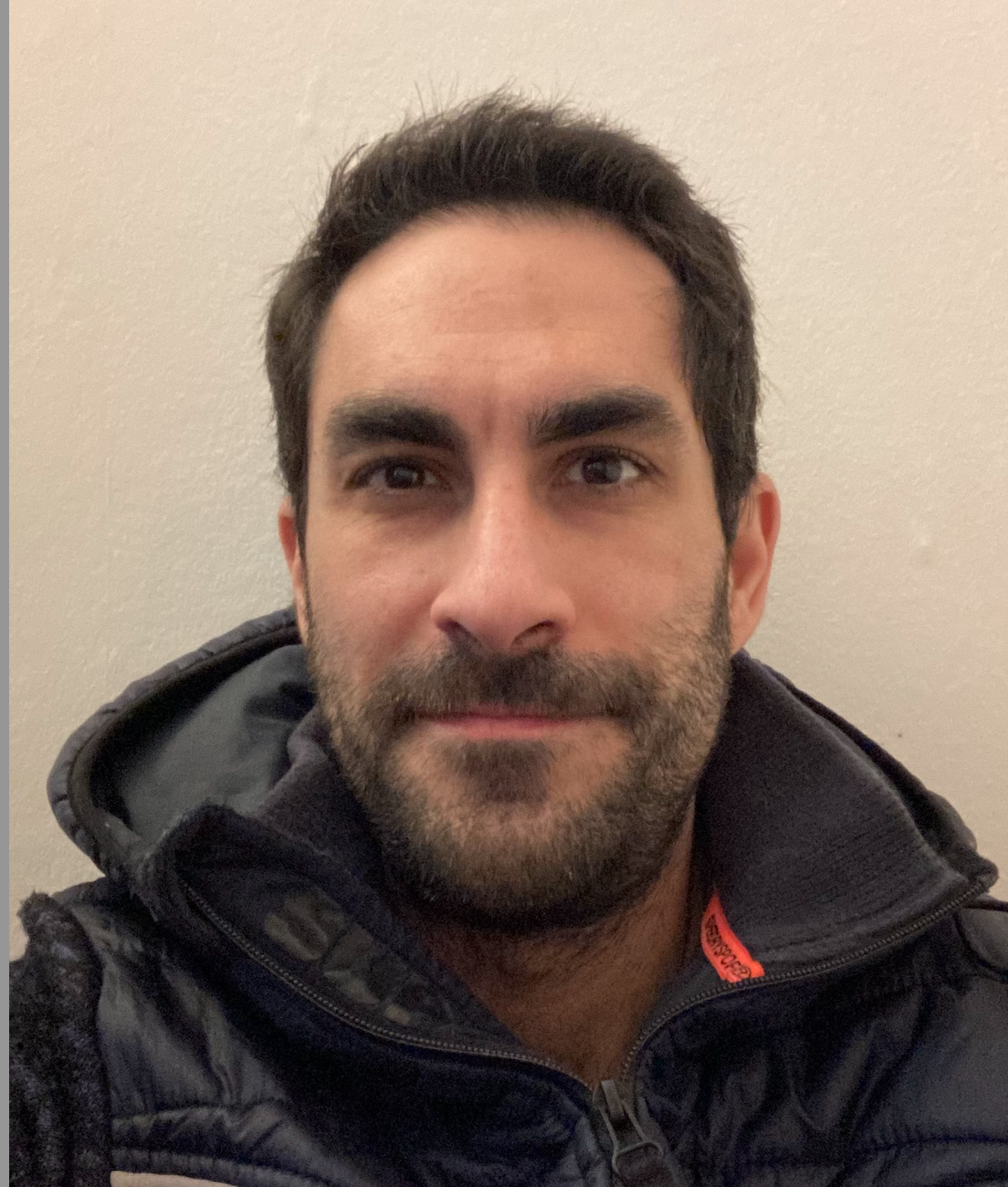}}]{Miguel Manuel de Villena} received the M.Sc. in energy engineering from the Technical University of Denmark in 2016 and the Ph.D. in electrical engineering from the University of Liège in 2021. He is currently a postdoctoral researcher in the Department of electrical engineering and computer science of the University of Liège. His research interests include the modelling and simulation of energy systems and local electricity markets and other energy markets.
\end{IEEEbiography}

\begin{IEEEbiography}[{\includegraphics[width=1in,height=1.25in,clip,keepaspectratio]{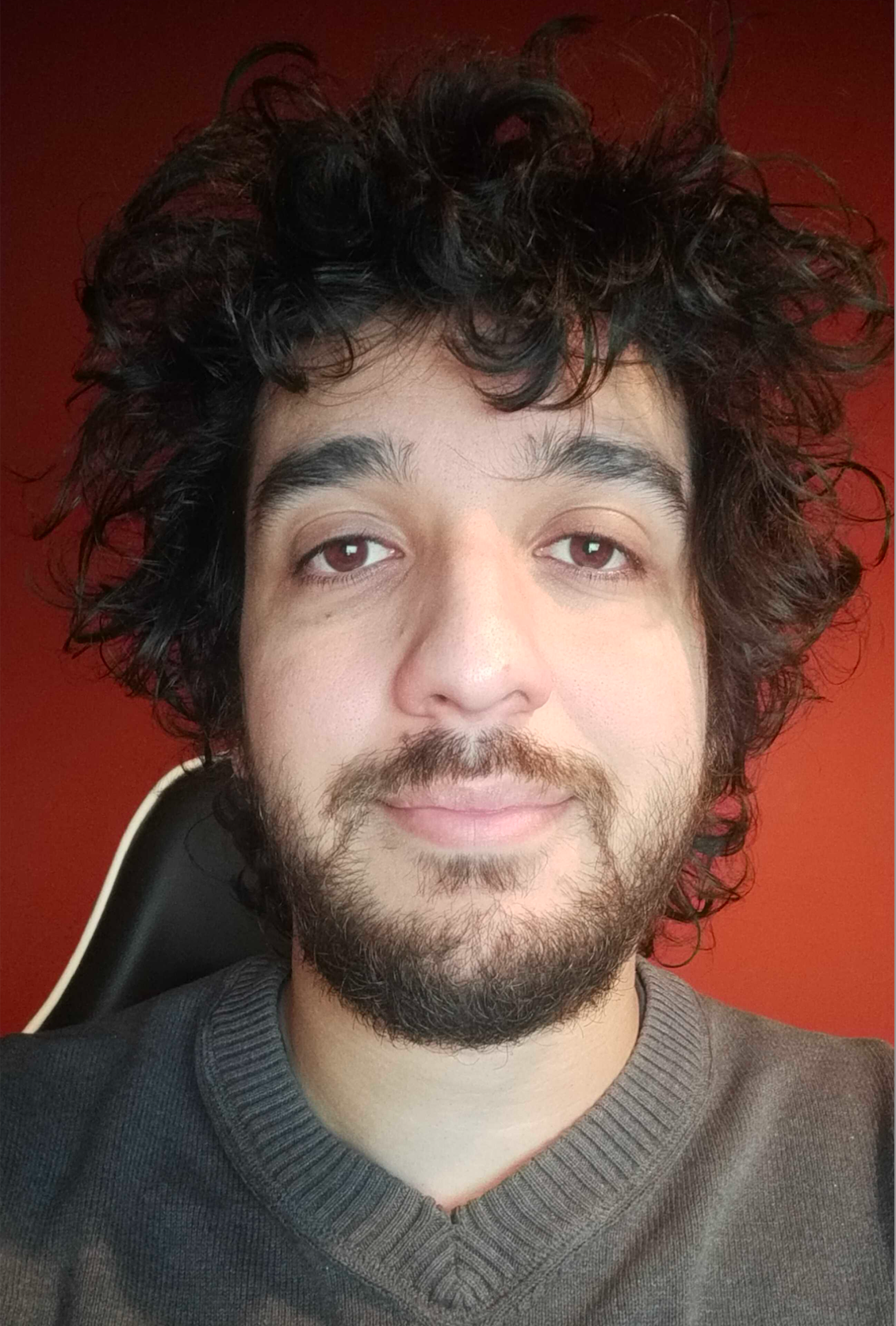}}] {Samy Aittahar} received the M.Sc. in computer science from the University of Rennes in 2015. He is currently a PhD student in the Department of electrical engineering and computer science of the University of Liège. His research interests includes the development and application of optimisation and reinforcement learning algorithms for energy system control tasks.
\end{IEEEbiography}

\begin{IEEEbiography}[{\includegraphics[width=1in,height=1.25in,clip,keepaspectratio]{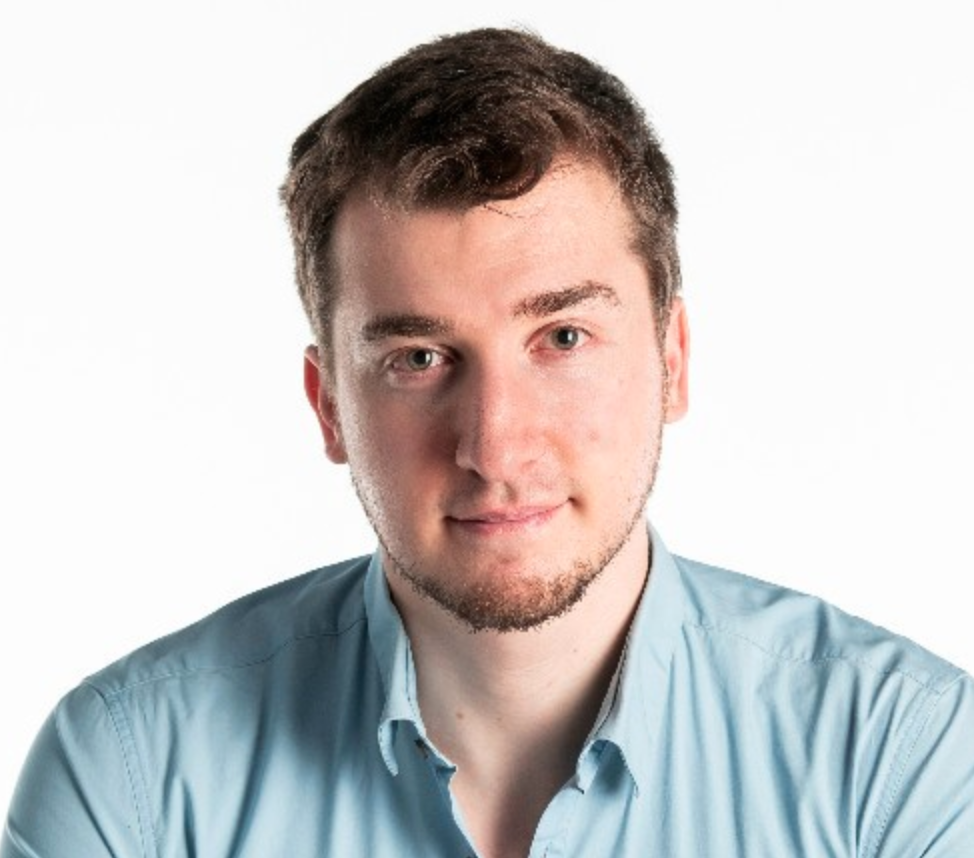}}] {Sébastien Mathieu} received its M.Sc. in electrical engineering in 2012 and its Ph.D. in 2016 from the University of Liège. His research interests are currently focused on electricity markets and energy audits.
\end{IEEEbiography}

\begin{IEEEbiography}[{\includegraphics[width=1in,height=1.25in,clip,keepaspectratio]{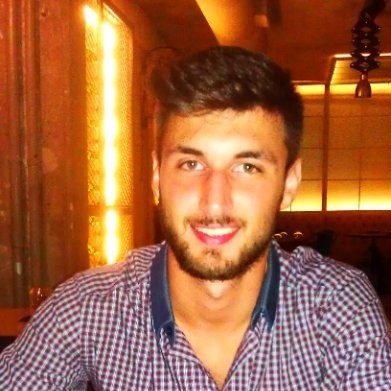}}]{Ioannis Boukas} obtained his Bsc. degree in mechanical engineering from the National Technical University of Athens, Greece and his Msc. in electro-mechanical engineering from the University of Liege, Belgium. He received his PhD on machine learning and optimization for energy storage management from the University of Liege, Belgium.
\end{IEEEbiography}

\begin{IEEEbiography}[{\includegraphics[width=1in,height=1.25in,clip,keepaspectratio]{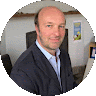}}]{Eric Vermeulen} received his Master of Arts in Economics form the University of Louvain-La-neuve in 1992. He then spends 2 years as a Research in Charge at the University and had several experience in the private sector. He is a founding member of the company haulogy in 2005, company developing intelligent softwares for the energy sector. Since 2019, he is in charge of the business development in the new energy markets: energy sharing, flexibility, e-mobility among others.
\end{IEEEbiography}

\begin{IEEEbiography}[{\includegraphics[width=1in,height=1.25in,clip,keepaspectratio]{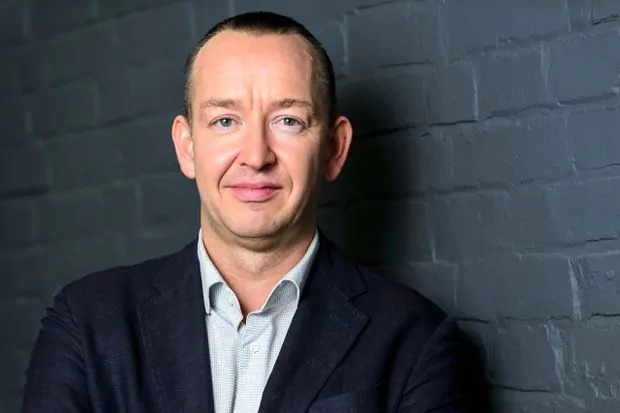}}]{Damien Ernst} received the M.Sc. and Ph.D. degrees in engineering from the University of Liège, Belgium, in 1998 and 2003, respectively. He is currently Full Professor at the University of Liège, and Visiting Professor at Télécom Paris. His research interests include electrical energy systems and reinforcement learning, a subfield of artificial intelligence.  He is also the CSO of Haulogy, a company developing intelligent software solutions for the energy sector. He has co-authored more than 300 research papers and two books. He has also won numerous awards for his research and, among which, the prestigious 2018 Blondel Medal. He is also regularly consulted by industries, governments, international agencies and the media for its deep understanding of the energy transition.
\end{IEEEbiography}

\EOD

\end{document}